\documentclass[preprint,12pt]{elsarticle}
\usepackage{amssymb}

\journal{Nuclear Physics A}
\begin{document}

\begin{frontmatter}



\title{Dynamics of two-cluster systems in phase space}


\author{Yu.A. Lashko\corref{cor1}}
\ead{ylashko@google.com} \cortext[cor1]{}
\author{G.F. Filippov}
\author{V.S. Vasilevsky}

\address{Bogolyubov Institute for Theoretical Physics, \\14-b Metrolohichna str., 03680, Kiev, Ukraine}

\begin{abstract}
We present a phase-space representation of quantum state vectors for
two-cluster systems. Density distributions in the Fock--Bargmann space
are constructed for bound and resonance states of $^{6,7}$Li and $^{7,8}$Be,
provided that all these nuclei are treated within a microscopic two-cluster
model. The density distribution in the phase space is compared with those
in the coordinate and momentum representations.  Bound states realize
themselves in a compact area of the phase space, as also do narrow
resonance states. We establish the quantitative boundaries of this region
in the phase space for the nuclei under consideration. Quantum trajectories
are demonstrated to approach their classical limit with increasing energy.
\end{abstract}


\end{frontmatter}



\section{Introduction}

\label{intro}

The idea of formulation of quantum mechanics in a phase space is discussed in
numerous theoretical papers \cite{1986JPSJ...55..762T, PhaseSpace2005,  1999JMP....40.2531M,  1993JChPh..98.3103T, %
1997JChPh.106.7228M, 1990JChPh..93.8862T}. The majority of such
investigations are concentrated on establishing the link between quantum and
classical mechanics. Due to the uncertainty principle there is no unique
definition of the phase space. By this reason, different quantum phase space
distribution have been proposed. In particular, in the phase-space
representation of Wigner and Husimi a quantum state is represented by a
distribution function (see the definitions, for instance, in \cite%
{1986JPSJ...55..762T, PhaseSpace2005}), and the equations of motion
are of the Liouville type.

A state-vector representation is another possibility to describe the
dynamics of a quantum system in phase space. In this case a quantum state is
represented by a wave function, and the equations of motion are of the Schr%
\"{o}dinger type. The definition of the phase-space representation is
related to the choice of an operator which should be diagonal in this
representation. In the coordinate representation the coordinate operator is
diagonal, while the momentum operator is non-local. At the same time, the
momentum representation diagonalizes the momentum operator, and the
coordinate operator is non-local. Obviously, the coordinate operator and the
momentum operator can not be diagonal simultaneously due to the uncertainty
principle. Hence, one should seek for another operator.

In Ref. \cite{1999JMP....40.2531M}, mention was made that representation of a
quantum state as a probability amplitude depending on two real variables
related to the coordinate and momentum dates back to the papers of Fock %
\cite{1928ZPhy...49..339F} and Bargmann \cite{Bargmann:1946me}. In the
Fock-Bargmann space, a quantum state is represented as an entire function of a
complex variable, with real and imaginary part of this variable being
proportional to the coordinate and momentum, correspondingly. The
Fock-Bargmann representation diagonalizes the creation operator, while the
annihilation operator is non-local.

A quantum problem can be resolved in any one of the above-mentioned
representations. The Fock-Bargmann image of a wave function can be
obtained from the wave function in the coordinate representation by a linear
mapping, while the Husimi and the Wigner distribution functions are bilinear
with respect to the wave function in the coordinate representation. However,
the Fock-Bargmann representation is closely related to the Husimi
distribution. The latter distribution is equal to the square of the
Fock-Bargmann image of the corresponding wave function multiplied by the
Bargmann measure.

In Refs. \cite{1993JChPh..98.3103T, 1997JChPh.106.7228M}, Torres-Vega
and Frederic suggested a quantum-state vector phase-space representation.
The authors postulated the existence of a complete basis of states $%
|q,p\rangle $ such that in the phase space a quantum state $|\psi \rangle $
is represented by an $\mathcal{L}^{2}(2)$ wave function $\psi (p,q)=\langle
q,p|\psi \rangle .$ Here $q,p$ are real values and the operators of
coordinate and momentum in this basis take the form:
\begin{equation}
\widehat{Q}={\frac{q}{2}}+\imath \hbar {\frac{\partial }{\partial p}},~~~%
\widehat{P}={\frac{p}{2}}-\imath \hbar {\frac{\partial }{\partial q}}.
\label{operator}
\end{equation}%
Thus Torres-Vega and Frederic developed a wave-function formulation of quantum
mechanics in a phase space. Wave functions are governed by the Schr\"{o}dinger
equation in the phase space, while square of absolute value of the wave
function plays the role of the probability density in the phase space
obeying the Liouville equation. However, the quantum-state vector
phase-space representation is not uniquely defined, because there exists an
infinite number of bases depending on two real variables $p,q$ which result
in the foregoing expression of the coordinate and momentum operators $%
\widehat{Q}$ and $\widehat{P}$.

In Ref. \cite{1997JChPh.106.7228M}, the authors used coherent states as basis
vectors and demonstrated that any coherent state representation leads to
expression (\ref{operator}) for the coordinate and momentum operators $%
\widehat{Q}$ and $\widehat{P}$. Moreover, they concluded that only the
coherent state representation makes it possible to define the operators of
coordinate and momentum in such a manner.

Following Klauder and Perelomov, in \cite{1997JChPh.106.7228M} a set of
coherent states is defined as a result of action of the Weyl operator $%
\widehat{D}(q,p)$ (a translation operator in the phase space)
\[
\widehat{D}(q,p)=\exp \left\{ {\frac{i}{\hbar }}\left( p\,\widehat{Q}-q%
\widehat{P}\right) \right\}
\]%
to any normalized vector $|\chi \rangle :$
\[
|q,p;\chi \rangle \equiv \widehat{D}(q,p)|\chi \rangle .
\]%
For any fixed vector $|\chi \rangle $ the set of coherent states $|q,p;\chi
\rangle $ ensures a continuous representation of quantum states where the
expansion coefficients can be interpreted as the wave function in the phase
space.

Then $|\psi_\chi(q,p)|^2$ is a "dilute" probability density in the phase space
that equals the probability for a system to be localized in some "dilute"
neighborhood of the center of the displaced state $\chi$. The degree of
"diffusiveness" depends on the choice of vector $\chi$. If we choose the
ground state of a harmonic oscillator as the vector $\chi$ then $|\psi_\chi(q,p)|^2 $ is probability to find a system in the elementary phase volume $\Delta q\Delta p=h$ near the point $(q,p)$.

The Schr\"{o}dinger equation in the phase space does not depend explicitly
on the vector $\chi .$ However, in calculating mean values of operators the wave
functions used for averaging should belong to the same coherent state
representation, i.e., to the same vector $\chi $. This could be ensured by
the requirement for the vector $\chi $ to be an eigenvector of a certain
operator. Hence, to formulate quantum mechanics unambiguously, one should
solve two equations in the phase space.

It is possible to formulate quantum mechanics in a phase space so that one
equation could be sufficient both for the solution of a quantum problem and
for specification of the representation. As concluded in \cite%
{1997JChPh.106.7228M}, the Fock-Bargmann representation suggests the best
answer to this question. The Fock-Bargmann representation is a
state-vector representation in the complex plane which is based on the
mapping of the pair of bosonic creation and annihilation operators $(
\widehat{a},\widehat{a}^{\dag }):$
\begin{equation}
\widehat{a}\rightarrow {\frac{\partial }{\partial R}},\,\,\,{\widehat{a}}%
^{\dag }\rightarrow R.  \label{boson_map}
\end{equation}%
In this representation quantum mechanics can be uniquely determined due to
the fact that the creation operator $\widehat{a}^{\dag }$ is diagonal in
this representation. The relation $[\widehat{a},\widehat{a}^{\dag }]=1$ is
followed by the relationships between the bosonic operators and the
coordinate and momentum operators $\widehat{Q}$ and $\widehat{P}$:
\[
\widehat{a}={\frac{1}{\sqrt{2}}}\left( {\frac{1}{b}}\widehat{Q}-i{\frac{b}{%
\hbar }}\widehat{P}\right) ,\,\,\widehat{a}^{\dag }={\frac{1}{\sqrt{2}}}%
\left( {\frac{1}{b}}\widehat{Q}+i{\frac{b}{\hbar }}\widehat{P}\right) ,
\]%
where $b$ is the oscillator length.

For a specific hamiltonian $\widehat{H}(\widehat{Q},\widehat{P})$ one can
fix the relationship between the operators $\widehat{Q}$, $\widehat{P}$ and
bosonic operators $(\widehat{a},\widehat{a}^{\dag })$ by choosing the value
of the oscillator length $b$. Then the Hamiltonian can be expressed in terms of
the bosonic operators $\widehat{H}(\widehat{Q},\widehat{P})\rightarrow \mathcal{H}%
(\widehat{a},\widehat{a}^{\dag })$. Hence, the vector $\chi $ is completely
determined by the value of $b$. In this case the vector $\chi $ is the vacuum
vector $|0\rangle $.

The Fock-Bargmann representation can be associated with any Glauber coherent
state representation based on the eigenvector $\chi $ of the annihilation
operator $a.$ According to Perelomov \cite{Perelomov_GCS}, a coherent
state describes a non-spreadable wave packet for an oscillator. Besides, coherent
states minimize the Heisenberg uncertainty relation $\Delta q\Delta p\geq h.$
Hence the coherent states are quantum states which resemble classic states the
most. The Schr\"{o}dinger equation in the Fock-Bargmann representation has
the form:
\begin{equation}
i\hbar {\frac{\partial }{\partial t}}|\psi \rangle =\mathcal{H}(\widehat{a},%
\widehat{a}^{\dag })|\psi \rangle \rightarrow i\hbar {\frac{\partial }{%
\partial t}}\psi _{FB}(R)=\mathcal{H}\left( {\frac{\partial }{\partial R}}%
,R\right) \psi _{FB}(R),  \label{eq:EqFB}
\end{equation}%
where
\[
R={\frac{1}{\sqrt{2}}}\left( {\frac{q}{b}}+i{\frac{b}{\hbar }}p\right) .
\]%
Hence, a complete basis of the wave functions belonging to the class of
coherent states which minimize the uncertainty relation can be unambiguously
determined as a result of solution of the Schr\"{o}dinger equation in the
Fock-Bargmann representation (\ref{boson_map}).

We follow precisely this strategy and present the wave functions and the
probability distributions in the Fock-Bargmann representation. However, we don't solve an equation of the type (\ref{eq:EqFB}), we use other
way for obtaining an exact wave function of a quantum system. We concentrate our
attention on the analysis of nuclear systems with a pronounced two-cluster
structure, whereas authors of Refs. \cite{1999JMP....40.2531M, %
1997JChPh.106.7228M} have studied one-dimensional systems wherein the
solutions could be found analytically. Much prominence is given to the
determination of those regions of phase space, which are the most important for
the dynamics of two-cluster systems. We also suggested the way of
investigation of the density distributions in the phase space in a
three-dimensional case, when the probability density distribution depends on
six variables: absolute values of coordinate and momentum and four angles.

There are some similarities between our approach and the Antisymmetrized
Molecular Dynamics (AMD) \cite{2003CRPhy...4..497K, %
2012PTEP.2012aA202K} and the Fermionic Molecular Dynamics (FMD) \cite%
{2000RvMP...72..655F, 2004NuPhA.738..357N}. The AMD and FMD have been
intensively employed to study a cluster structure of atomic nuclei and both of
them appeal to a phase space. All three methods make use the same single
particle orbitals to construct a many-particle wave function in the form of
the Slater determinant. It means that all methods involve the same part of
the total Hilbert space, provided that all of them take into account the
same partition (or clusterization) of $A$ nucleon system. However, there are
more differences between our approach and AMD\ and FMD. First, we use the
Slater determinants as the generation functions for a complete basis of
many-particle oscillator functions, describing the most important, from
physical point of view, types of motion of many-particle system. In our
approach cluster parameters are the generator coordinates that allow us to
select necessary basis functions from the infinite set of oscillator
functions. Meanwhile, in the AMD\ and FMD they are independent variables in
the phase space. By using a set of many-particle oscillator functions, we
reduce the Schr\"{o}dinger equation to the matrix form. The AMD and FMD make
use of the time-dependent equations derived from the time-dependent variational
principle. Having obtained the wave function in a discrete, oscillator
representation, we then transform it to the Fock-Bargmann or phase space,
where we analyze phase trajectories of a quantum many-particle system.

It should be emphasized that all the results presented in this paper are
obtained within the phase-space formulation of quantum mechanics which is
valid both for finite $\hbar $ and when $\hbar $ goes to zero. We consider
some simple model cases as well as two-cluster systems. The model cases,
such as a three-dimensional harmonic oscillator and a plane wave, help us to
reveal peculiarities of phase space portraits. However, our main aim is to
study phase portraits of real physical systems, namely, light atomic nuclei.
All calculations are performed within a microscopic two-cluster model based
on the resonating-group method (RGM) \cite{1937PhRv...52.1083W}. Our
analysis started from assumptions of the following cluster structure of the
nuclei under consideration: $^{6}Li=\alpha +d$, $^{7}Li=\alpha +t$, $%
^{7}Be=\alpha +^{3}He$, $^{8}Be=\alpha +\alpha $. Instead of solving an
integro-differential equation in a phase space as the authors of Ref. \cite%
{1993JChPh..98.3103T} do, we deal with the RGM Hamiltonian in the
representation of the Pauli-allowed harmonic-oscillator states defined in the
Fock--Bargmann space. Doing so, we reduce the integro-differential equation
to a set of linear equations for coefficients of the expansion of the wave
function in the harmonic-oscillator basis.

Our paper is organized as follows. In Section \ref{Sec:FockBargmann} we
introduce all necessary definitions and formulate an approach for
constructing density distributions in the Fock-Bargmann space. In Section \ \ref%
{Sec:ModelProblems}, the effectiveness of the suggested approach is demonstrated
for two simple model problems: a harmonic oscillator and a free motion in 3D
space. In Section \ref{Sect:2ClModel} we give a brief review of the employed two-cluster
model. Details of the calculations are shown in Section \ref{Sect:Details}.
Phase portraits for bound and resonance states in the light atomic nuclei are
presented in Section \ref{Sect:Results}. Finally, in Section \ref%
{Sect:Conclusion} we conclude the main results obtained.

\section{The Fock-Bargmann representation \label{Sec:FockBargmann}}

In this section  and Section \ref{Sec:ModelProblems}  we shall use dimensionless units for energy and
length. The energy $E$ is measured in units of $\frac{\hbar^{2}}{mb^{2}}$ and the length is
measured in terms of oscillator length $b.$ The value of the oscillator length will be determined in what follows.

The transition from the wave function in the coordinate or momentum
representation to the wave function in the Fock--Bargmann representation is
performed using the Bargmann-Segal integral transformation
\begin{equation}
\Psi _{E}(\mathbf{R})=\int K(\mathbf{R},\mathbf{r})\Psi _{E}(\mathbf{r})d%
\mathbf{r},~~\mathbf{R}={\frac{\vec{\xi}+i\vec{\eta}}{\sqrt{2}}},
\label{eq:01}
\end{equation}%
the kernel of which is the modified Bloch-Brink orbitals \cite%
{2005PPN..36.714F}:
\begin{equation}
K(\mathbf{R},\mathbf{r})={\frac{1}{{\pi ^{3/4}}}}\exp \left( -{\frac{r^{2}}{2%
}}+\sqrt{2}\,(\mathbf{rR})-{\frac{R^{2}}{2}}\right) .  \label{eq:02}
\end{equation}%
In the Fock-Bargmann space, wave functions are entire analytical functions of
complex variable $\mathbf{R}$, while $\vec{\xi}$ and $\vec{\eta}$ are coordinate and momentum vectors, respectively.

The modified Bloch---Brink orbital is an eigenfunction of the coordinate
operator in the Fock-Bargmann space:
\[
\widehat{\mathbf{r}}K(\mathbf{R},\mathbf{r})=\mathbf{r}K(\mathbf{R},\mathbf{r%
}),~~~~\widehat{\mathbf{r}}={\frac{1}{\sqrt{2}}}\left( \mathbf{R}+\nabla _{%
\mathbf{R}}\right) .
\]%
On the other hand, the modified Bloch-Brink orbital is a coherent state and
generates a complete basis of functions of a 3D-harmonic oscillator.

Having calculated the wave function in the Fock-Bargmann space, we can find
the density distribution, which depends on six variables: absolute values of the
coordinate and momentum $\xi$, $\eta $ and four angles $\Omega_{\vec{\xi}}$%
, $\Omega_{\vec{\eta}}$:
\begin{equation}
dD _{E}(\vec{\xi},\vec{\eta})=|\Psi _{E}(R)|^{2}d\mu _{B}=|\Psi _{E}(\mathbf{%
R})|^{2}\exp \left( {-\frac{\eta ^{2}+\xi ^{2}}{2}}\right) {\frac{d\vec{\xi}d%
\vec{\eta}}{(2\pi )^{3}}}.  \label{dro_def}
\end{equation}%
Here
\[
d\mu _{B}=\exp \{-(\mathbf{R}\cdot \mathbf{S})\}{\frac{d\vec{{\xi }}d\vec{{%
\eta }}}{(2\pi )^{3}}},~~\mathbf{S}=\mathbf{R}^{\ast }
\]%
is the Bargmann measure.

In essence, $dD _{E}(\vec{\xi},\vec{\eta})$ is a density matrix in the
Fock-Bargmann space. It is positively defined for all values of $\vec{\xi}$
and $\vec{\eta}$ as opposed to the Wigner function, which is often used to
construct a density distribution in a phase space. In fact, $D _{E}(\vec{\xi},%
\vec{\eta})$ coincides with the Husimi distribution within definition of
variables\ $\vec{\xi}$ and $\vec{\eta}$.

It is rather difficult to investigate the density distribution $D_{E}(\vec{\xi},%
\vec{\eta}),$ because it depends on six variables. To solve this problem and
extract as much as possible physical information, we propose to integrate $%
D_{E}(\vec{\xi},\vec{\eta})$ over the solid angles $\Omega _{\vec{\xi}}$ and $%
\Omega _{\vec{\eta}}$.
Then we come out with the density
distribution $D_{E}(\xi ,\eta )$ which depends only on the absolute values
of the coordinate and momentum.
\[
D_{E}(\xi ,\eta )\equiv {\frac{\int d\Omega _{\vec{\xi}}\int d\Omega _{\vec{%
\eta}}\,dD_{E}(\vec{\xi},\vec{\eta})}{d\xi d\eta }}.
\]%
It is important to recall, that the variable $\xi $ is analogue of the distance
between interacting particles (clusters) and the variable $\eta $ represents a
momentum of relative motion of particles (clusters).

Given the energy $E$ and other integrals of motion, like the orbital
momentum and parity, the density distribution $D_E(\xi,\eta)$ is comprised
of the infinite number of phase trajectories. The phase trajectories are
determined as a continuous set of points in the $(\xi ,\eta )$ plane for the
fixed values of the density distribution $D_E(\xi ,\eta )=const$, while the
probability of realization of the phase trajectory is proportional to the
values of $D_E(\xi,\eta)$. Hereinafter we shall call an infinite set of phase
trajectories characterizing a particular quantum state of the system under
study the phase portrait.

With increase in energy $E$,
all the quantum phase trajectories gradually
approach their classical limit. Therefore, analyzing phase portraits of
quantum systems we can make a quantitative estimate of the energy such that
the maximum of the density distribution $D_{E}(\xi ,\eta )$ falls on the
classical trajectory.

In \cite{2013FBS.55.817L}, we have constructed and analyzed the phase
portraits for the free motion of a 1D quantum particle and for the motion in
the field of Gaussian potential. In the present paper we concentrate on the
three-dimensional case. Before proceeding to the analysis of phase portraits of
two-cluster systems, let us discuss the density distribution for the states
of a three-dimensional oscillator with the number of quanta $N$ and for a free
motion of a 3D-particle in the states with the orbital momentum $l.$

\section{Model problems\label{Sec:ModelProblems}}

\subsection{Harmonic oscillator}

The overlap integral of the modified Bloch-Brink orbitals
\[
\langle \mathbf{S}|\mathbf{R}\rangle \equiv \int K(\mathbf{S},\mathbf{r})K(%
\mathbf{R},\mathbf{r})d\mathbf{r}=\exp (\mathbf{R}\mathbf{S})
\]%
generates a complete basis for the harmonic-oscillator functions in the
Fock-Bargmann representation:
\begin{equation}
\exp (\mathbf{R}\mathbf{S})=\sum_{n=0}^{\infty
}\sum_{l}N_{nl}^{2}R^{2n+l}S^{2n+l}\sum_{m}Y_{lm}(\Omega _{R})Y_{lm}^{\ast
}(\Omega _{S})  \label{eq:03}
\end{equation}%
Here both the absolute value and the solid angle of the complex vector $%
\mathbf{R}$ take complex values:
\[
R=\sqrt{\mathbf{R}^{2}}=\sqrt{\frac{1}{2}}\sqrt{\left( {\xi ^{2}-\eta ^{2}}%
\right) {+2i(\vec{\xi}\vec{\eta})}},\,\,\,\,\Omega _{R}={\frac{\mathbf{R}}{R}%
}={\frac{\vec{\xi}+i\vec{\eta}}{\sqrt{\left( \xi ^{2}-\eta ^{2}\right) +2i(%
\vec{\xi}\vec{\eta})}}}
\]

Each basis function corresponds to the total number of quanta $N=2n+l,$
where $l$ is the orbital momentum and $n$ is the number of radial quanta:
\begin{equation}
|n,l,m;\mathbf{R}\rangle =N_{nl}R^{2n+l}Y_{lm}(\Omega _{R}),\,\,\,N_{nl}=%
\sqrt{{\frac{4\pi }{(2n)!!(2n+2l+1)!!}}}.  \label{eq:04}
\end{equation}%
Hence, according to the definition (\ref{dro_def}), a density distribution
for the $|n,l;\mathbf{R}\rangle $ state of harmonic oscillator should be
written as
\[
dD _{n, l}(\vec{\xi},\vec{\eta})={\frac{1}{2l+1}}\sum_{m}|n,l,m;\mathbf{S}%
\rangle \langle n,l,m;\mathbf{R}|\,d\mu _{B}.
\]%
Here averaging over projection $m$ of the orbital momentum is performed.

\begin{figure}[tbh]
\begin{center}
\includegraphics[width=\textwidth]{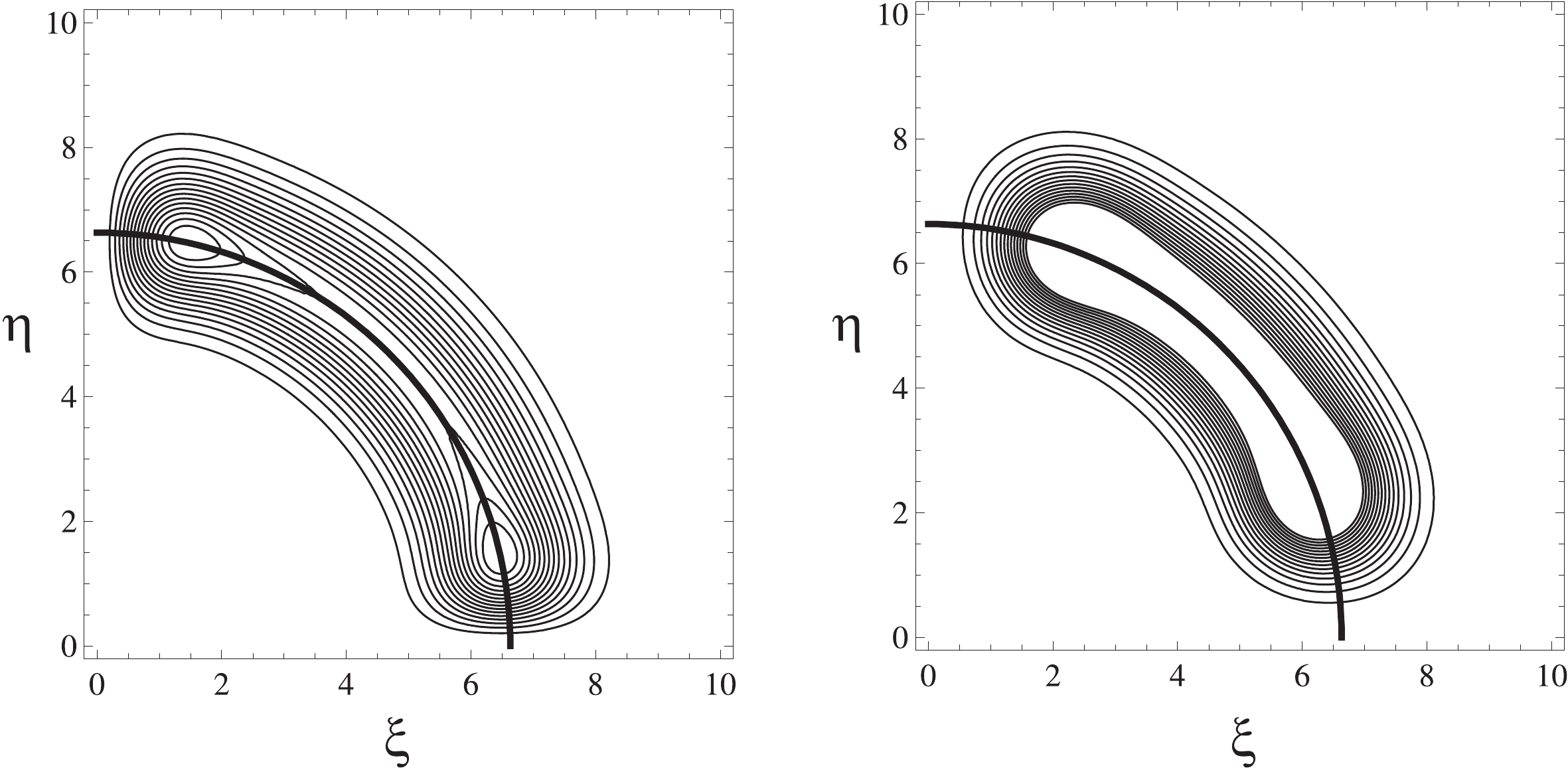}
\end{center}
\par
\caption{Phase portraits of a 3D oscillator $D _{n, l}(\protect\xi ,%
\protect\eta )=const$ with the total number of quanta $2n+l=20$ as a
function of dimensionless coordinate $\protect\xi $ and momentum $\protect%
\eta $. The left panel corresponds to $D _{n=10, l=0}(\protect\xi ,\protect%
\eta )=const,$ the right panel demonstrates $D _{n=5, l=10}(\protect\xi ,\protect%
\eta )=const$. The heavy line on both panels denotes the classical trajectory defined by Eq. (\ref{real_max_l}).}
\label{fig:1}
\end{figure}

Integrating over the solid angles related to the coordinate $\vec{\xi}$ and momentum
$\vec{\eta}$, we come to the expression
\begin{eqnarray*}
D _{n,l}(\xi ,\eta ) &=&{\frac{\xi ^{2}\eta ^{2}}{(2\pi )^{3}}}\int
d\Omega _{\xi }\int d\Omega _{\eta }dD _{n,l}(\vec{\xi},\vec{\eta})={%
\frac{\xi ^{2}\eta ^{2}}{2\pi ^{2}}}\,{\frac{N_{nl}^{2}}{2^{2n+l}}}\,e^{-{%
(\xi ^{2}+\eta ^{2})/2}}\times \\
&\times &\sum_{\nu =0}^{\left[ {\frac{l}{2}}\right] }d_{\nu }^{l}(\xi
^{2}-\eta ^{2})^{2n+2\nu }(\xi ^{2}+\eta ^{2})^{l-2\nu } \\
& \times & \,_{2}F_{1}\left( -n-\nu ,{\frac{1}{2}},{\frac{3}{2}};-{\frac{%
4(\xi \eta )^{2}}{(\xi ^{2}-\eta ^{2})^{2}}}\right) ,
\end{eqnarray*}%
where
\[
d_{\nu }^{l}={\frac{(-1)^{\nu }(2l-2\nu )!}{2^{l}\nu !(l-\nu )!(l-2\nu )!}}.
\]%
As observed in the right panel of Fig. \ref{fig:1}, with increase in the orbital momentum $l$ the
density distribution $D _{n, l}(\xi ,\eta )$ is slightly forced out from
the region of small $\xi $ and $\eta$ and narrows.

The left panel of Fig. \ref{fig:1} demonstrates that the density distributions $D
_{n, l}(\xi ,\eta )$ corresponding to the same total number of quanta $N$, but
different orbital momenta $l,$ peak on the same circle:
\begin{equation}
{\frac{\xi ^{2}+\eta ^{2}}{2}}=2n+l+2.  \label{real_max_l}
\end{equation}%
Eq. (\ref{real_max_l}) describes a classical trajectory of the particle with the energy $E_{cl}=2n+l+2.$
Hence, the most probable locus of points in the phase space for the eigenstate
of a 3D harmonic oscillator with the number of quanta $N=2n+l$ corresponds
to the classical energy $E_{cl}=N+2$ rather than to $E_{cl}=N+3/2.$ The maximum
of density distribution for the $N$th state of a 1D harmonic oscillator
was matched by the classical energy $E_{cl}=N$ instead of $E_{cl}=N+1/2$
(see \cite{1993JChPh..98.3103T, 2013FBS.55.817L}).

\subsection{Plane wave\label{Sec:PlaneWave}}

In the Fock-Bargmann representation, a plane wave corresponding to the
momentum $\mathbf{k}$ becomes:
\begin{equation}
\phi _{\mathbf{k}}(\mathbf{R})=\pi ^{-3/4}\exp \left\{ -{\frac{k^{2}}{2}}-i%
\sqrt{2}(\mathbf{kR})+{\frac{R^{2}}{2}}\right\} .  \label{plane_wave}
\end{equation}%
It is an eigenfunction of the momentum operator $\widehat{\mathbf{k}}$:
\[
\widehat{\mathbf{k}}\phi _{\mathbf{k}}(\mathbf{R})=\mathbf{k}\phi _{\mathbf{k%
}}(\mathbf{R}),~~~~\widehat{\mathbf{k}}=-{\frac{i}{\sqrt{2}}}\left( \mathbf{R%
}-\nabla _{\mathbf{R}}\right) .
\]%
The density distribution for the plane wave (\ref{plane_wave}) has the Gaussian
dependence on the momentum $\eta $ and is unaffected by the coordinate $\xi
: $
\begin{equation}
dD _{\mathbf{k}}(\vec{\xi},\vec{\eta})=|\phi _{\mathbf{k}}(\mathbf{R}%
)|^{2}d\mu _{B}=\pi ^{-3/2}\exp \left( -(\mathbf{k}-{\vec{\eta}})^{2}\right)
{\frac{d\vec{\xi}d\vec{\eta}}{(2\pi )^{3}}}.  \label{dro_plane}
\end{equation}%
Obviously, the density distribution (\ref{dro_plane}) peaks on the line ${%
\vec{\eta}}=\mathbf{k}$. This line coincides with the classical phase trajectory
of a free particle with momentum $\mathbf{k}.$

The problem becomes more intricate if we consider free motion of a
particle in the states with given value of the orbital momentum $l$. We can
expand the eigenfunction (\ref{plane_wave}) of the momentum operator $\widehat{k}$ in the
basis of the harmonic oscillator functions (\ref{eq:04}) in the Fock-Bargmann
space:
\begin{equation}
\phi _{\mathbf{k}}(\mathbf{R})=\sum_{n=0}^{\infty }\sum_{l}\sum_{m}|n,l,m;%
\mathbf{R}\rangle C_{nl}(k)Y_{lm}(\Omega _{k}).  \label{plane_wave_exp}
\end{equation}%
Expansion (\ref{plane_wave_exp}) in the Fock-Bargmann representation is much
the same as the multipole expansion of the plane wave in the coordinate
representation.

The expansion coefficients $C_{nl}(k)$ coincide with the basis functions of
harmonic oscillator in the momentum representation. In particular, for zero orbital momentum the expansion
coefficients $C_{nl}(k)$ are defined as
\begin{equation}
C_{n,l=0}(k)=\sqrt{{\frac{2\Gamma (n+1)}{\Gamma (n+3/2)}}}%
L_{n}^{1/2}(k^{2})\exp \left( -{\frac{k^{2}}{2}}\right) .  \label{coef_pl_w}
\end{equation}%
Generally in the
Fock-Bargmann space any wave function $\Psi _{lm}^{E}(\mathbf{R})$ of a quantum state characterized by the orbital momentum $%
l, $ its projection $m$ and energy $E$ can be represented as the expansion
into the basis of the harmonic oscillator functions (\ref{eq:04}):
\begin{equation}
\Psi _{lm}^{E}(\mathbf{R})=\sum_{n=0}^{\infty }\sum_{l}C_{nl}^{E}|n,l,m;%
\mathbf{R}\rangle ,  \label{eq:05}
\end{equation}%
where the expansion coefficients $C_{nl}^{E}$ are the solutions of a set of
linear algebraic equations to which the Schr\"{o}dinger equation is reduced.
The expansion (\ref{eq:05}) is written in the Fock-Bargmann space; it can
be also presented in coordinate and momentum spaces with the same set of the
expansion coefficients $\left\{ C_{nl}^{E}\right\} $. Such way of
representing and calculating the wave functions is the essence
of the method we follow in the present paper.

Hence, the density distribution for the quantum state described by the wave
function $\Psi _{lm}^{E}(\mathbf{R})$ takes the form:
\[
dD _{E, l}(\vec{\xi},\vec{\eta})={\frac{1}{2l+1}}\sum_{n=0}^{\infty }\sum_{%
\tilde{n}=0}^{\infty }\sum_{m}C_{nl}^{E}C_{\tilde{n}l}^{E}|k,l,m;\mathbf{S}%
\rangle \langle k,l,m;\mathbf{R}|\,d\mu _{B}.
\]%
Upon integrating the foregoing expression over the solid angles $\Omega_{\vec{\xi}}$
and $\Omega_{\vec{\eta}}$ we obtain:
\begin{eqnarray*}
D _{E, l} (\xi ,\eta ) &=&{\frac{\xi \eta }{2\pi ^{2}}}(\xi ^{2}-\eta
^{2})\,\exp \left( -{\frac{\xi ^{2}+\eta ^{2}}{2}}\right)
\,\sum_{n=0}^{\infty }\sum_{\tilde{n}=0}^{\infty }N_{nl}N_{\tilde{n}%
l}C_{nl}^{E}C_{\tilde{n}l}^{E}\times \\
&\times &\sum_{\nu =0}^{\left[ {\frac{l}{2}}\right] }d_{\nu }^{l}\left( {%
\frac{\xi ^{2}+\eta ^{2}}{2}}\right) ^{l-2\nu }(\xi ^{2}-\eta ^{2})^{n+%
\tilde{n}+2\nu } \\
& \times & Im  \left[ B\left( {\frac{1}{2}}+i{\frac{\xi \eta }{\xi ^{2}-\eta
^{2}}};n+\nu +1,\tilde{n}+\nu +1\right) \right] ,
\end{eqnarray*}%
where $B(z;a,b)$ is the incomplete beta function.

\begin{figure}[tbh]
\begin{center}
\includegraphics[width=\textwidth]{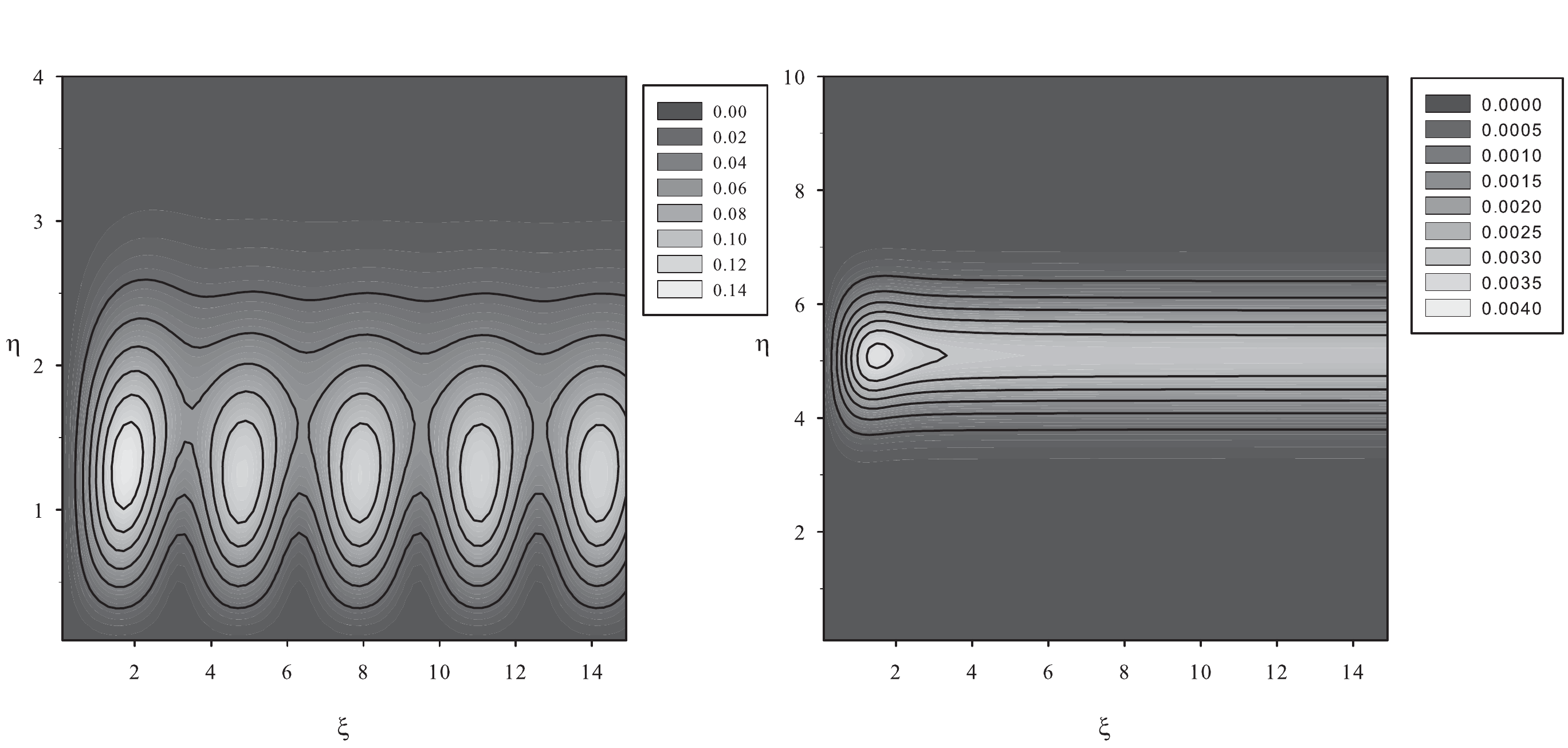}
\end{center}
\par
\caption{Phase portraits of a free 3D particle with the energy $E=k^{2}/2$ and
zero orbital momentum as a function of the dimensionless coordinate $\protect\xi
$ and momentum $\protect\eta $. The left panel corresponds to $k=1$, while
the right panel corresponds to $k=5.$}
\label{fig:3}
\end{figure}

Substituting the coefficients (\ref{coef_pl_w}) in the formula given above we
derive the density distribution for a free motion of the particle with zero
orbital momentum. Figure \ref{fig:3} demonstrates the phase portraits for a free
particle with $k=1$ and $k=5.$
As observed in Fig. \ref{fig:3}, for $k=1$ the density distribution
oscillates in the variable $\xi $. These oscillations, resulting in a number of closed trajectories,
are of quantum nature.
With increasing the energy the density distribution becomes smoother; all the
quantum trajectories become infinite and approach the classical trajectory $\eta \simeq k.$

\begin{figure}[tbh]
\begin{center}
\includegraphics[width=\textwidth]{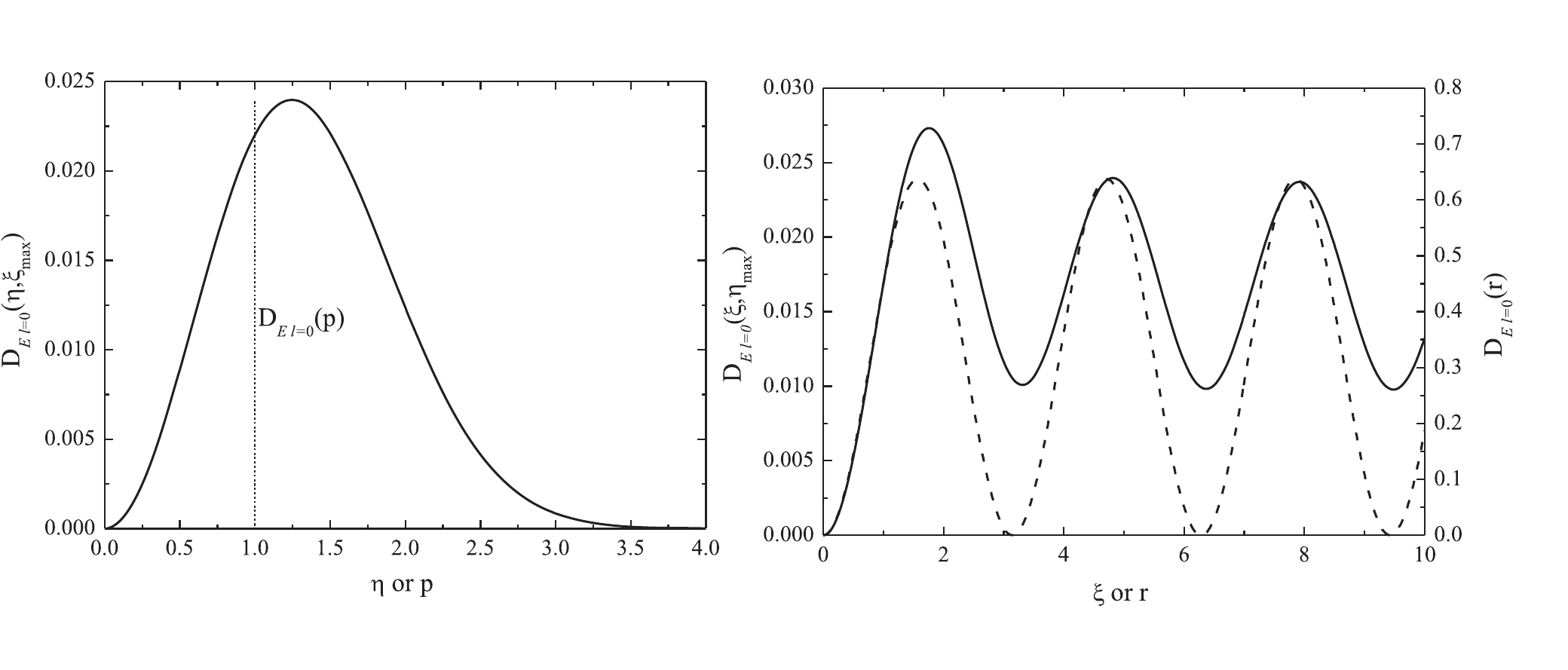}
\end{center}
\par
\caption{The left panel presents the density distributions $D _{E, l=0}(\protect\xi _{max},\protect\eta )$ in the phase space (a solid line) and $D _{E, l=0}(p)$ in the momentum space (a dashed line) for a free 3D particle with the energy $E=k^{2}/2,$ $k=1$ and zero orbital momentum. The right panel shows the density distributions $D _{E, l=0}(\protect\xi ,\protect\eta _{max})$ in the phase space (a solid line) and $D _{E, l=0}(p)$ in the coordinate space (a dashed line).}
\label{fig:4}
\end{figure}

Figure \ref{fig:4} compares the density distributions in the Fock-Bargmann
space, coordinate representation and momentum representation for a free 3D
particle with the energy $E=k^{2}/2$, $k=1$ and zero orbital momentum.

In the coordinate representation the wave function of a free particle in the state
with $l=0$ is described by the spherical Bessel function $\psi _{k}\left(
r\right) =\sqrt{2/\pi }j_{0}(kr),$ while in the momentum representation it
is just the Dirac delta-function $\psi _{k}\left( p\right) =\delta (p-k)$.
As is clear from Fig. \ref{fig:4}, locations of the maxima for the density
distributions in the variable $\xi $ in the Fock-Bargmann and coordinate
representations almost coincide, while the density distribution in the
variable $\eta $ peaks at $\eta \simeq 1.25$ instead of $\eta =1$ as
the momentum density distribution does.

\section{Two-cluster model of light nuclei \label{Sect:2ClModel}}

We present shortly main ideas of the two-cluster model that we are going to
use to study the dynamics of two-cluster systems. First of all, we restrict
ourselves with the lightest nuclei of $p$-shell.

A wave function for the two-cluster partition $A=A_{1}+A_{2}$ is represented as
\begin{equation}
\Psi_{J}=\widehat{\mathcal{A}}\left\{ \left[ \Phi_{1}\left( A_{1}\right)
\Phi_{2}\left( A_{2}\right) \right] _{S}\psi_{LS}^{J}\left( q\right)
Y_{L}\left( \widehat{\mathbf{q}}\right) \right\} _{J},  \label{eq:C01}
\end{equation}
where $\widehat{\mathcal{A}}$ is the antisymmetrization operator, 
$\mathbf{q}$ is the Jacobi vector which is proportional to the
vector $\mathbf{r} $ connecting the centers of mass of the
interacting clusters
\begin{equation}
\mathbf{q}=\mathbf{r}\sqrt{\frac{A_{1}A_{2}}{A_{1}+A_{2}}}=\sqrt{\frac {%
A_{1}A_{2}}{A_{1}+A_{2}}}\left[ \frac{1}{A_{1}}\sum_{i\in A_{1}}\mathbf{r}%
_{i}-\frac{1}{A_{2}}\sum_{j\in A_{2}}\mathbf{r}_{j}\right] .  \label{eq:C02}
\end{equation}
It is assumed that the number of nucleons in each cluster $A_{i}$ does not
exceed 4: 1$\leq A_{i}\leq$4\ . As one sees, we use the $LS$ scheme of
coupling, when
the total spin $S$ is coupled with the total orbital momentum $L$ and generates
the total angular momentum $J$. For two interacting $s$-clusters, the total
orbital momentum $L $ coincides with the angular momentum of the relative
motion of clusters.

To find the wave function $\psi_{LS}^{J}\left( q\right) $ of relative motion of
the clusters one has to solve a set of integro-differential equations (see the details
in Ref. \cite{kn:wilderm_eng}). The set of equations and the form of the wave
function (\ref{eq:C01}) are the key elements of the well known Resonating
Group Method.

An integral part of the integro-differential equations for the functions $%
\psi_{LS}^{J}\left( q\right) $  is due the to the antisymmetrization
operator $\widehat{\mathcal{A}}$ and originates from the potential, and kinetic
energy operators, and the norm kernel. The latter means that interaction
between clusters is energy dependent. By neglecting the operator $\widehat {%
\mathcal{A}}$, we obtain differential equations with a local potential. This
approximation is called the folding approximation or the folding model.

The Form and procedure of solving the equations for the functions $\psi _{LS}^{J}\left(
q\right) $ can be simplified, if we use a set of square-integrable
functions. Such way of solving the equations of the Resonating Group Method with
oscillator functions is called an Algebraic Version of the Resonating Group
Method \cite{kn:Fil_Okhr, kn:Fil81}. To realize the algebraic
version, we introduce a set of cluster oscillator functions%
\begin{equation}
\left\vert n,L,S,J\right\rangle =\widehat{\mathcal{A}}\left\{ \left[ \Phi
_{1}\left( A_{1}\right) \Phi _{2}\left( A_{2}\right) \right] _{S}\Phi
_{nL}\left( q\right) Y_{L}\left( \widehat{\mathbf{q}}\right) \right\} _{J},
\label{eq:C10}
\end{equation}%
where%
\begin{eqnarray}
\Phi _{nL}\left( r,b\right) &=&\left( -1\right) ^{n}\mathcal{N}%
_{nL}~b^{-3/2}\rho ^{L}e^{-\frac{1}{2}\rho ^{2}}L_{n}^{L+1/2}\left( \rho
^{2}\right) ,\quad  \label{eq:C12} \\
\rho &=&\frac{r}{b},\quad \mathcal{N}_{nL}=\sqrt{\frac{2\Gamma \left(
n+1\right) }{\Gamma \left( n+L+3/2\right) }}  \nonumber
\end{eqnarray}%
is an oscillator function in the coordinate space ($b$ is the oscillator length).
We are interested in wave functions and density distributions both in
the coordinate and in momentum space. They will be constructed with the
help of the oscillator functions in the momentum space%
\begin{eqnarray}
\Phi _{nL}\left( p,b\right) &=&\mathcal{N}_{nL}~b^{3/2}\rho ^{L}e^{-\frac{1}{%
2}\rho ^{2}}L_{n}^{L+1/2}\left( \rho ^{2}\right) ,  \label{eq:C13} \\
\quad \rho &=&p\cdot b,  \nonumber
\end{eqnarray}%
The cluster oscillator functions $\left\vert n,L,S,J\right\rangle $ are totally
antisymmetric and compose a complete set of basis functions with specific
physical properties. Namely, they are a part of the total Hilbert space
describing the $A_{1}+A_{2}$ clusterization of the system of $A$ nucleons with
fixed internal clusters functions $\Phi _{1}\left( A_{1}\right) $ and $\Phi
_{2}\left( A_{2}\right) $. The cluster oscillator functions can be used to
expand any wave function of the type (\ref{eq:C01}):%
\begin{equation}
\Psi _{J}=\sum_{LS}\sum_{n=0}^{\infty }C_{nL}\left\vert n,L,S,J\right\rangle
.  \label{eq:C15}
\end{equation}%
The expansion coefficients $\left\{ C_{nL}\right\} $ represent a two-cluster wave
function in a discrete, oscillator representation and obey the system of
linear equations%
\begin{equation}
\sum_{LS}\sum_{m=0}^{\infty }\left[ \left\langle n,L,S,J\left\vert \widehat{H%
}\right\vert m,L,S,J\right\rangle -E\cdot \left\langle
n,L,S,J|m,L,S,J\right\rangle \right] C_{mL}=0,  \label{eq:C16}
\end{equation}%
where $\left\langle n,L,S,J\left\vert \widehat{H}\right\vert
m,L,S,J\right\rangle $ is a matrix element of the hamiltonian between the cluster
oscillator functions and $\left\langle n,L,S,J|m,L,S,J\right\rangle $ is an
overlap of these functions or the norm kernel. For two $s$-clusters, the
norm kernel has a very simple form%
\begin{equation}
\left\langle n,L,S,J|m,L,S,J\right\rangle =\delta _{n,m}\lambda _{n}.
\label{eq:C16A}
\end{equation}%
The constants $\lambda _{n}$ are the eigenvalues of the antisymmetrization operator.
The states $\left\vert n,L,S,J\right\rangle $ with $\lambda _{n}=0$ are called
the Pauli-forbidden states. The antisymmetric basis functions with the
quantum numbers $n,L,S,J$ do not participate in constructing the wave
function (\ref{eq:C01}). Only the Pauli-allowed states (i.e., the states with $%
\lambda _{n}>0$) take part in describing the dynamics of the two-cluster system.

Equation (\ref{eq:C16A}) indicates that the cluster oscillator functions (%
\ref{eq:C10}) are not normalized to unity despite the fact that the functions $%
\Phi _{1}\left( A_{1}\right) $, $\Phi_{2}\left( A_{2}\right) $, $\Phi
_{nL}\left( q\right) $ and $Y_{LM}\left( \widehat{\mathbf{q}}\right) $ are
properly normalized. The antisymmetrization operator $\widehat{\mathcal{A}}$
is responsible for that. By renormalizing the basis functions%
\[
\left\vert \overline{n,L,S,J}\right\rangle =\left\vert n,L,S,J\right\rangle /%
\sqrt{\lambda_{n}}
\]
and the expansion coefficients $\overline{C}_{nL}=C_{nL}\sqrt{\lambda_{n}}$, we
arrive at the standard matrix form of the Schr\"{o}dinger equation with the
orthonormal basis%
\begin{equation}
\sum_{LS}\sum_{m}\left[ \left\langle \overline{n,L,S,J}\left\vert \widehat {H%
}\right\vert \overline{m,L,S,J}\right\rangle -E\cdot\delta_{n,m}\right]
\overline{C}_{mL}=0,  \label{eq:C16B}
\end{equation}
where $n$ and $m$ numerates only the Pauli-allowed states.

Formally, the expansion (\ref{eq:C15}) of the wave function contains an infinite
set of basis functions; however, actually we need a large but restricted
set of functions. In an oscillator representation, situation is similar to
the coordinate form of the Schr\"{o}dinger equation where one needs to find a wave
function up to a finite distance $R_{a}.$ Beyond this point the well-known
asymptotic form of the wave function is valid. $R_{a}$ determine a distance where a short-range
interaction is negligibly small and an asymptotic part of the hamiltonian is
dominant. The same is true for the discrete representation. We need to
calculate the wave function up to the finite value of $n=N_{a}$; starting from
this number of quanta an asymptotic form for the expansion coefficients of the wave function is valid.
Like $R_{a}$, the parameter $N_{a}$ draw
a border between an internal and asymptotic regions. When solving
the Schr\"{o}dinger equation numerically both in the coordinate and oscillator
representations, the parameters $R_{a}$ and $N_{a}$ are used as variational
parameters. One has to determine minimum values of $R_{a}$ and $N_{a}$ such that
further increasing of them do not change the results of calculations.

To solve the system of equations (\ref{eq:C16}) or (\ref{eq:C16B}), one needs to
take into account the boundary conditions. The asymptotic form for a wave
function of a bound state in the coordinate space
\begin{eqnarray}
\psi_{LS}^{J}\left( q\right) & \sim & \exp\left\{ -\kappa q\right\} /q,
\label{eq:C20C} \\
\kappa & = & \sqrt{\frac{2m\left\vert E\right\vert }{\hbar^{2}}},  \nonumber
\end{eqnarray}
as was shown in Ref. \cite{kn:Fil81}, is transformed into%
\begin{eqnarray}
C_{nL} & \sim & \sqrt{R_{n}}\exp\left\{ -\kappa bR_{n}\right\} /R_{n},
\label{eq:C20O} \\
R_{n} & = & \sqrt{4n+2L+3}  \nonumber
\end{eqnarray}
for the expansion coefficients of the wave function. Similar relations \ are
valid for the wave function of a continuous spectrum state (%
\textit{single channel case)} in the coordinate space
\begin{eqnarray}
\psi_{LS}^{J}\left( q\right) & \sim & \sin\left( kq+\delta_{L}+L\frac{\pi}{2}%
\right) /q,  \label{eq:C21C} \\
k & =& \sqrt{\frac{2mE}{\hbar^{2}}}  \nonumber
\end{eqnarray}
and in the oscillator representation%
\begin{equation}
C_{nL}\sim \sqrt{R_{n}}\sin\left( kb\sqrt{R_{n}}+\delta_{L}+L\frac{\pi }{2}%
\right) /R_{n},  \label{eq:C21O}
\end{equation}
where $\delta_{L}$ is a phase shift. To avoid bulky formulae and long
additional explanations, in equations (\ref{eq:C20C})-(\ref{eq:C21O}) we
showed an asymptotic form of the wave functions for neutral clusters. Similar
expression can be written for charged clusters. 

There is an equivalent form for the expansion (\ref{eq:C15}). We can write
expansion for the intercluster wave function by using the same set of
the expansion coefficients%
\begin{equation}
\psi_{LS}^{J}\left( q\right) =\sum_{n=0}^{\infty}C_{nL} \Phi_{nL}\left(
r,b\right)  \label{eq:C17}
\end{equation}
Similar form can be used to determine intercluster function $%
\psi_{LS}^{J}\left( p\right) $ in momentum space. The functions $%
\psi_{LS}^{J}\left( q\right) $ and $\psi_{LS}^{J}\left( p\right) $ are
connected by the Fourier-Bessel transformation%
\begin{equation}
\psi_{LS}^{J}\left( p\right) =\sqrt{\frac{2}{\pi}}\int_{0}^{%
\infty}dqq^{2}j_{L}\left( pq\right) \psi_{LS}^{J}\left( q\right)
\label{eq:C18}
\end{equation}

One important note should be made. The wave function (\ref{eq:C01}) for bound
states is traditionally normalized to unity%
\[
\left\langle \Psi_{J}|\Psi_{J}\right\rangle =\sum_{n=0}^{\infty}\left\vert
\overline{C}_{nL}\right\vert ^{2}=1
\]
but it is not the case for the corresponding intercluster functions
\begin{equation}
\left\langle \psi_{LS}^{J}|\psi_{LS}^{J}\right\rangle =S_{LJ}.
\label{eq:C19}
\end{equation}
In the oscillator representation it reads as%
\begin{equation}
S_{LJ}=\sum_{n=0}^{\infty}\left\vert C_{nL}\right\vert
^{2}=\sum_{n=0}^{\infty}\left\vert \overline{C}_{nL}\right\vert
^{2}/\lambda_{n}.  \label{eq:C19A}
\end{equation}
The deviation of the quantity $S_{LJ}$ from unity shows how strong is the
effect of the Pauli principle.

Having calculated the expansion coefficients $\left\{ C_{nL}\right\} $, we
can easily construct the wave functions and density distributions in the coordinate,
momentum, and the Fock-Bargmann or phase spaces.

We do not dwell on the calculation of matrix elements of the hamiltonian between
the cluster oscillator functions (\ref{eq:C10}). We refer the reader to the review
\cite{kn:cohstate2E}, where all necessary formulae are presented. They could
help one to calculate the matrix elements of the hamiltonian, which include the central
and spin-orbital components of the nucleon-nucleon forces and the Coulomb
interaction as well.

\section{Details of calculations \label{Sect:Details}}

Our main objective is to study light nuclei with a pronounced two-cluster
structure. Among these nuclei are the \ $^{6}Li$, \ $^{7}Li$ and $^{7}Be,$
because the two-cluster decay thresholds $\alpha+d$, $\alpha+^{3}H$ and $%
\alpha+^{3}He$, respectively, lie not far from the ground state of the
nuclei and other two- and three-cluster thresholds are at higher energy.
There are the strong grounds to believe that the channels $\alpha+d$, $%
\alpha+^{3}H$ and $\alpha+^{3}He$ are responsible to a great extent for
the structure of bound and low-lying resonance states in $^{6}Li$, \ $^{7}Li$
and $^{7}Be$ nuclei. We also consider the $^{8}Be$ as $\alpha+\alpha$
configuration, which generates a set of the rotational $0^{+}$, $2^{+}$ and $%
4^{+}$ resonance states.

In our calculations we make use of the Minnesota nucleon-nucleon potential
suggested by Tang and coworkers. The central part of the potential is taken from
Ref. \cite{kn:Minn_pot1} and the spin-orbital components are taken from Ref.
\cite{1970NuPhA.158..529R} (IV version).

In such type of calculations we have got two free parameters. The first
parameter, the oscillator length $b$, we use to minimize the energy of the selected
two-body threshold. In other words, we use this parameter to optimize
description of the internal cluster structure. The second parameter $u,$ determining
the odd components of the Minnesota potential, is fitted to reproduce the bound
state energies for all nuclei but $^{8}Be$. The latter nucleus has no bound
state and the "ground state" actually is a very narrow resonance state (its
energy $E$=0.0918 MeV and width \ $\Gamma$=5.57 eV) that can be treated as
a quasi-stationary state. Thus for $^{8}Be$, we find the parameter $u$, which
reproduces fairly well the energy and width of the $0^{+}$ resonance state. To study
peculiarities of the $0^{+}$ resonance state in the $^{8}Be$, we make two different
calculations. One was mentioned above. This result we mark as a Resonance
State (RS).In the second variant of calculation, which will be marked as a Bound
State (BS), we switch off the Coulomb interaction and thus obtain a bound
state in $^{8}Be.$

To study effects of the Coulomb interaction in the mirror nuclei \ $^{7}Li$ and $%
^{7}Be$, we use the same input parameters, which were adjusted for the $^{7}Li$
ground state. In this case the difference in position of the bound and resonance
states in \ $^{7}Li$ and $^{7}Be$ is due to the Coulomb interaction.

In the present model, the total spin $S$ and the total orbital momentum $L$
are the quantum numbers. For the two-cluster configurations, which are taken into
considerations, the total spin is determined by the second cluster, as the
first cluster, alpha-particle, has the spin equals zero.

We use the following scheme of calculations. First, we construct matrix
elements of the hamiltonian and other operators of physical importance between
the cluster oscillator functions. We use $N_{a}$= 200 oscillator functions in all
our calculations. This number of basis functions provides us with convergent
and stable results both for the bound states and scattering states as well.
Second, we calculate the eigenvalues and eigenfunctions of the hamiltonian. As a
result we obtain a bound state (if any) and a large set of pseudo-bound states.
The latter are the states of continuous spectrum with wave functions
normalized to unity in a fixed basis of functions%
\[
\sum_{n=0}^{N_{a}-1}\left\vert C_{n}\left( E_{\alpha}\right) \right\vert ^{2}=1
\]
and obeying the conditions%
\[
C_{N_{a}}\left( E_{\alpha}\right) =0,
\]
where $E_{\alpha}$ is the energy of the $\alpha$th ($\alpha=$0, 1, 2, \ldots, $N_{a}$-1)
pseudo-bound state. By using the eigenfunctions $\left\{ C_{n}\left(
E_{\alpha}\right) \right\} $, we construct the density distributions in
the coordinate, momentum, and Fock-Bargmann spaces.

Third, we calculate the phase shifts of elastic scattering by solving a system
of linear equations, which takes into account the proper boundary conditions for
scattering states. It allows us to determine energy and width of resonance
states. 

In Table \ref{Tabl:InputParm} we show the input parameters of the calculations and
the spectrum of bound and resonant states of the light atomic nuclei. We also
indicate in the Table the dominant two-cluster channel taken into
consideration. The experimental data are from Refs. \cite{2002NuPhA.708....3T,
2004NuPhA.745..155T}. As we see, with the input parameters, indicated
in Table, we obtain a fairly good description of the bound and resonance states
comparing to the experimental data. However, the energy and width of few resonances
slightly differ from the experimental values. This can be attributed to the
peculiarities of the used nucleon-nucleon potential and restrictions of the
present model (using, for instance, the same oscillator length for both
interacting clusters). We consider this drawback of the present calculations
not crucial for the interpretation and validity of the results which will be
discussed bellow.

\begin{table}[htbp] \centering%
\caption{Spectrum of bound and resonance states of the light nuclei and the input parameters of
calculations. Calculated energy and width are given in MeV, the experimental values of energy
and width are given in MeV$\pm$keV. The theoretical and experimental width of the $0^+$ resonance state in $^8Be$ are presented in eV.}%
\begin{tabular}{|l|l|l|l|l|l|l|l|l|}
\hline
\multicolumn{2}{|l}{System} & \multicolumn{2}{|l}{Input} &  &
\multicolumn{2}{|l}{Theory} & \multicolumn{2}{|l|}{Experiment} \\ \hline
Nucleus & $A_{1}+A_{2}$ & $b$, fm & $u$ & $J^{\pi}$ & $E$ & $\Gamma$ & $E$ &
$\Gamma$ \\ \hline
$^{6}Li$ & $\alpha+d$ & 1.3110 & 0.9254 & $1^{+}$ & -1.4750 & - & -1.4743 & -
\\
&  &  &  & $3^{+}$ & 0.8480 & 0.0284 & 0.712 $\pm$ 2 & 0.024 $\pm$ 2 \\
&  &  &  & $2^{+}$ & 4.2880 & 3.0052 & 2.838 $\pm$ 22 & 1.30 $\pm$ 100 \\
\hline
$^{7}Li$ & $\alpha+^{3}H$ & 1.3451 & 0.969 & $\frac{3}{2}^{-}$ & -2.4676 & -
& -2.4670 & - \\
&  &  &  & $\frac{1}{2}^{-}$ & -1.6040 & - & -1.9894 & - \\
&  &  &  & $\frac{7}{2}^{-}$ & 2.4710 & 0.1285 & 2.185 & 0.069 \\
&  &  &  & $\frac{5}{2}^{-}$ & 4.9390 & 1.7121 & 4.137 & 0.918 \\ \hline
$^{7}Be$ & $\alpha+^{3}He$ & 1.3451 & 0.969 & $\frac{3}{2}^{-}$ & -1.6302 & -
& -1.5866 & - \\
&  &  &  & $\frac{1}{2}^{-}$ & -0.8161 & - & -1.1575 & - \\
&  &  &  & $\frac{7}{2}^{-}$ & 3.3360 & 0.2232 & 2.98 $\pm$ 50 & 0.175 $\pm$
7 \\
&  &  &  & $\frac{5}{2}^{-}$ & 5.7420 & 2.0207 & 5.14 $\pm$ 100 & 1.2 \\
\hline
$^{8}Be$ & $\alpha+\alpha$ & 1.3736 & 0.950 & $0^{+}$ & 0.0818 & 2.40 &
0.0918 & 5.57 $\pm$ 0.25 \\
&  &  &  & $2^{+}$ & 1.2840 & 0.6418 & 3.12 $\pm$ 10 & 1.513 $\pm$ 15 \\
&  &  &  & $4^{+}$ & 9.7970 & 3.5827 & 11.44 $\pm$ 150 & $\approx$ 3.500 \\
\hline
\end{tabular}
\label{Tabl:InputParm}%
\end{table}%

\section{Results and discussion \label{Sect:Results}}

\begin{figure}[ptbh]
\begin{center}
\includegraphics[width=\textwidth]{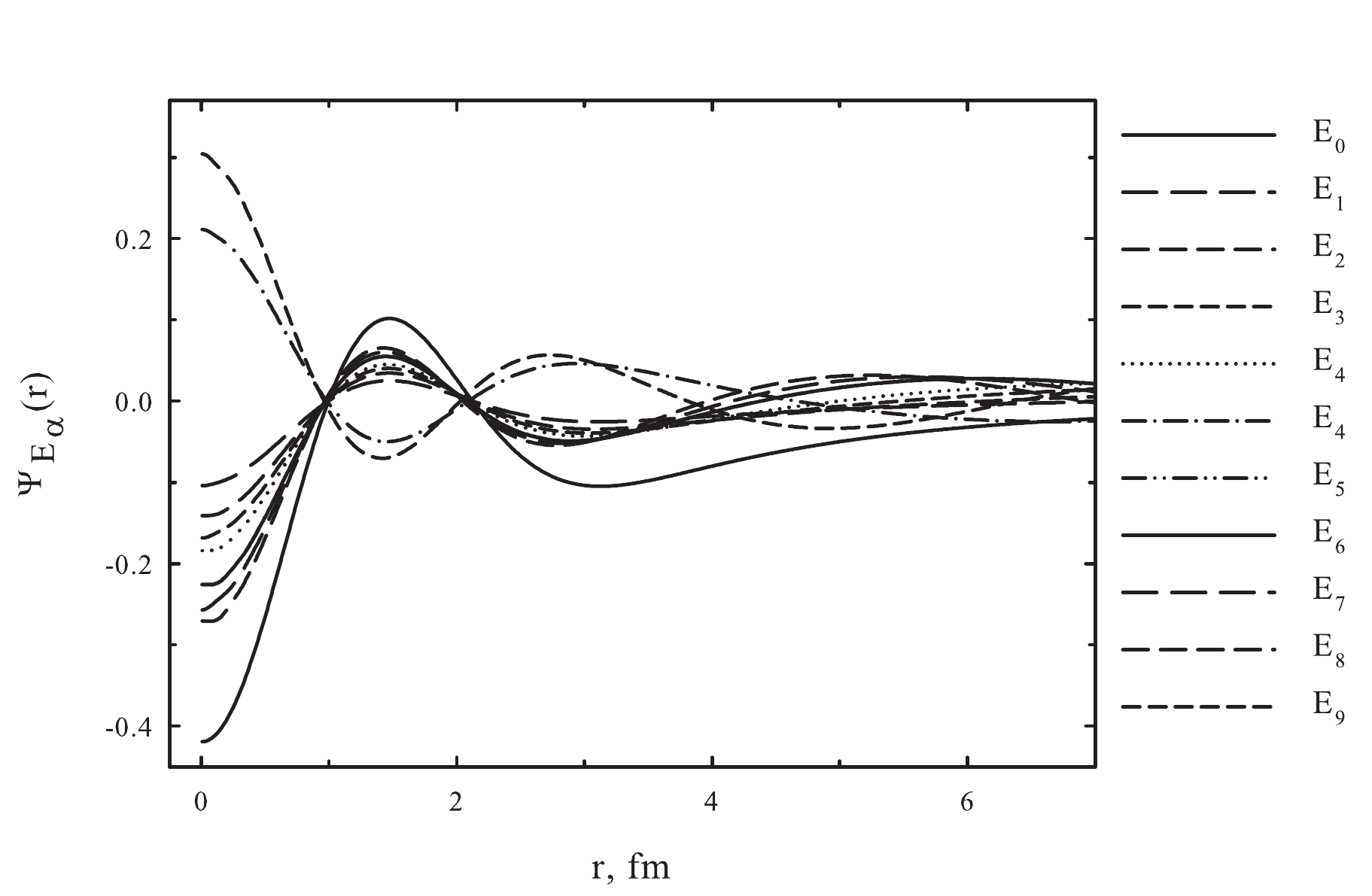}

\end{center}
\caption{Coordinate wave functions of 10 lowest $0^{+}$ states in $^{8}Be$.}
\label{Fig:WavFuns108BeCS}
\end{figure}

We start our discussions with wave functions in the coordinate space. In Figure %
\ref{Fig:WavFuns108BeCS} we present the wave functions of the $^{8}Be$ for 10 lowest
$0^{+}$ eigenstates of two-cluster hamiltonian. One notices that all the wave
functions have nodes approximately at the same points in the coordinate space.
This feature is typical for all nuclei under consideration. Position and the
number of nodes depends on the nucleus and its angular momentum. \ These fixed
nodes results from the orthogonality of the wave function $\psi_{LS}^{J}$ of
intercluster motion to the wave functions of the forbidden states; and the number
of nodes is equal to the number of the Pauli-forbidden states. This fact is
the key element of the Orthogonality Condition Model \cite{Saito69, %
kn:Saito77}, which is a simplified version of the Resonating Group Method
with an approximate treatment of the Pauli principle and a local cluster-cluster
interaction.

The wave functions in the momentum space have also nodes, but their structure is
not so simple and pictorial.

\begin{figure}[ptbh]
\begin{center}
\includegraphics[width=\textwidth]{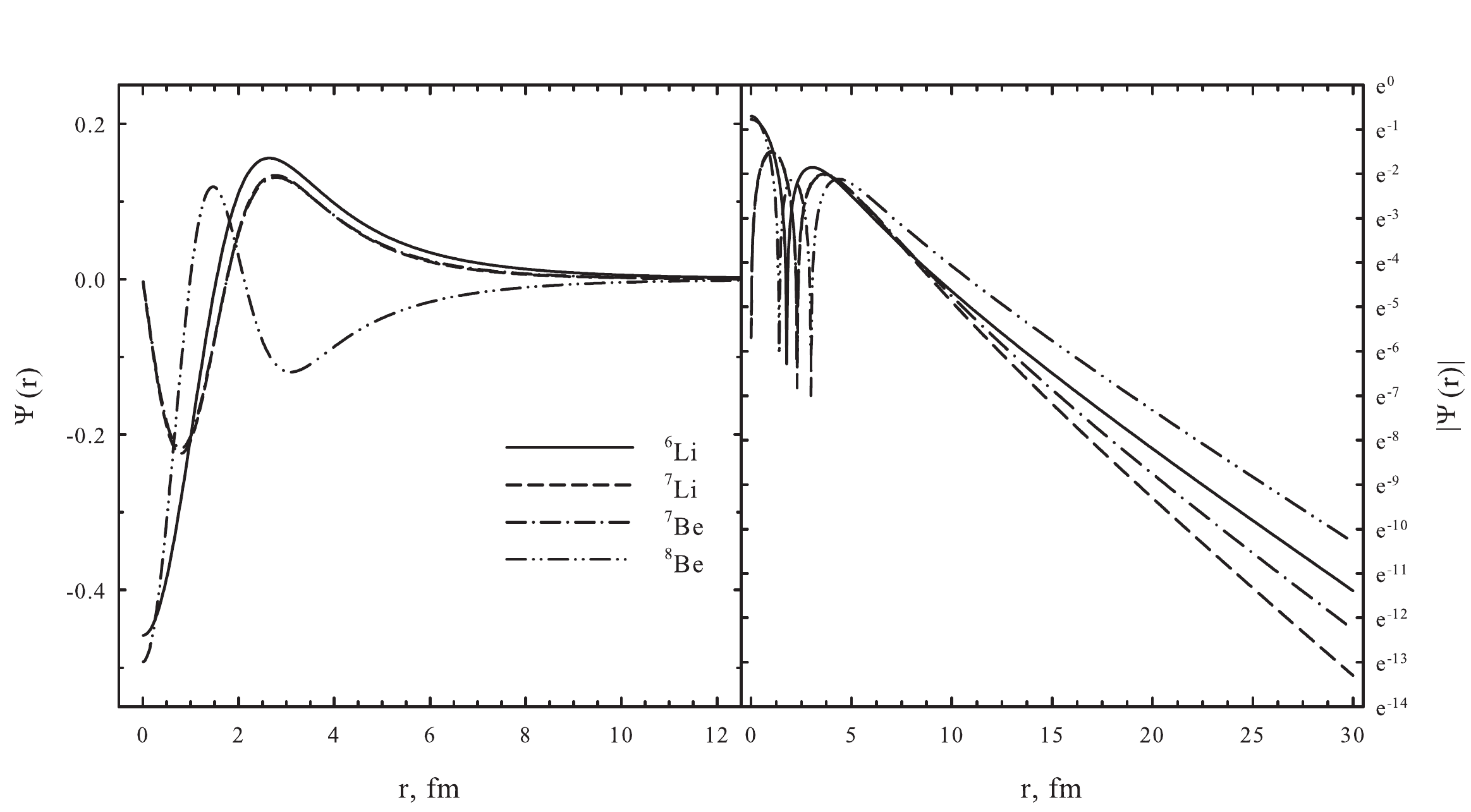}
\end{center}
\caption{Wave functions of the ground state of $^{6}Li$,
$^{7}Li$, $^{7}Be$ and $^{8}Be$ in coordinate space. A short range behavior is shown on the
left-hand side, while an asymptotic behavior is demonstrated on the right-hand
side of the Figure.}
\label{Fig:WavFunsGS}
\end{figure}

In Figure \ref{Fig:WavFunsGS} we compare the coordinate wave functions of the
ground states in $^{6}Li$, $^{7}Li$, $^{7}Be$ and $^{8}Be$ (BS). The wave
functions of $^{7}Li$ and $^{7}Be$ are indistinguishable in this Figure,
despite that the energy difference of the bound states is 0.84 MeV.

To demonstrate whether our basis of cluster functions is large enough to
provide correct results, in Figure \ref{Fig:WavFunsGS} we also display
an asymptotic behavior of the wave functions. One can see that the wave functions are
decreased as $\exp\left\{ -\kappa q\right\} /q$ at large values of $q$. The
order of the curves depends on the bound state energy: the smaller is
the energy, the lower is the corresponding curve.


\subsection{Phase portrait of bound states}

\begin{figure}[ptbh]
\begin{center}
\includegraphics[width=\textwidth]{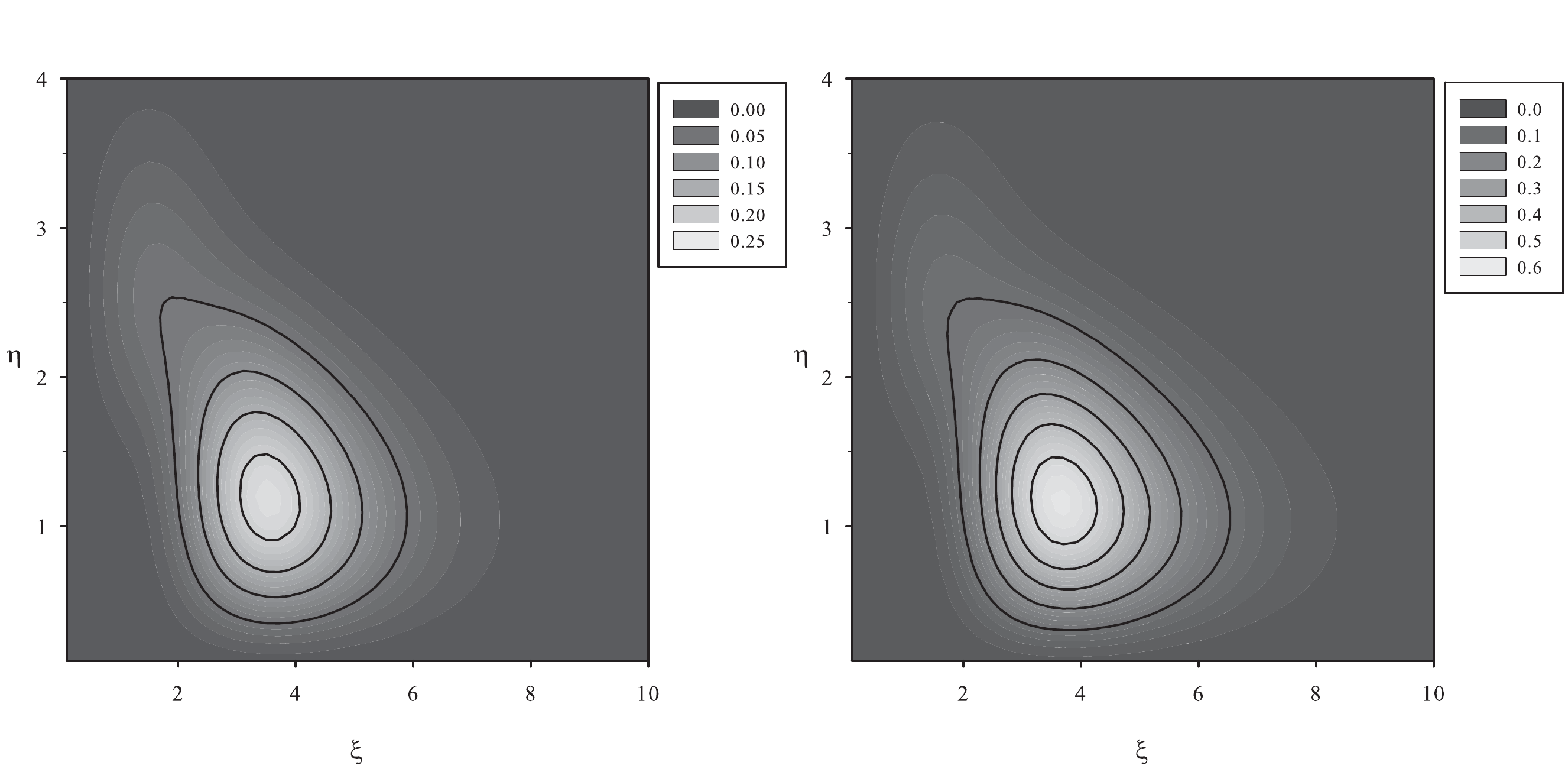}
\end{center}
\caption{Phase portraits of the $3/2^{-}$ (left) and $1/2^{-}$ (right) bound
states in $^{7}Li$.}
\label{Fig:PP_BStates7Li}
\end{figure}

Now we turn our attention to the phase portraits of bound states of
the two-cluster systems. \ In Figures \ref{Fig:PP_BStates7Li}, \ref%
{Fig:PP_BStates7Be} we display the phase portraits of the $^{7}Li$ and $^{7}Be$
bound states. The phase portrait for the $^{6}Li$ ground state was shown in our
previous publication \cite{2013FBS.55.817L}. The general feature of these
figures is that the two-cluster system is concentrated in a rather narrow region
of the phase space, despite the fact that some bound states are weakly bound
ones. As was expected, the more dispersed is a state in the coordinate space, the
more compact it is in the momentum space. And vise versa. It is interesting to
note that a maximum of the density distributions in the phase space of the bound states
lies at $\eta \approx 1.0$ and $3.5\leq \xi \leq 4$. The above-mentioned
value of $\eta $ is noticeable differs from the dimensionless momentum $k$
(which for the deepest bound state in $^{7}Li$ equals $k$=0.46). In our
opinion, this indicates that the shape of the density distribution for a bound
state has a pure quantum character.

As we see from Fig. \ref{Fig:WavFunsGS}, the wave functions and thus the density
distributions of the ground states in the coordinate space have nodes. There are also
nodes in the momentum space for a wave function of bound states. However, the
density distribution of the states in the phase space does not have any node
in the range $0<\eta < \infty$ and $0<\xi < \infty$.

\begin{figure}[ptbh]
\begin{center}
\includegraphics[width=\textwidth]{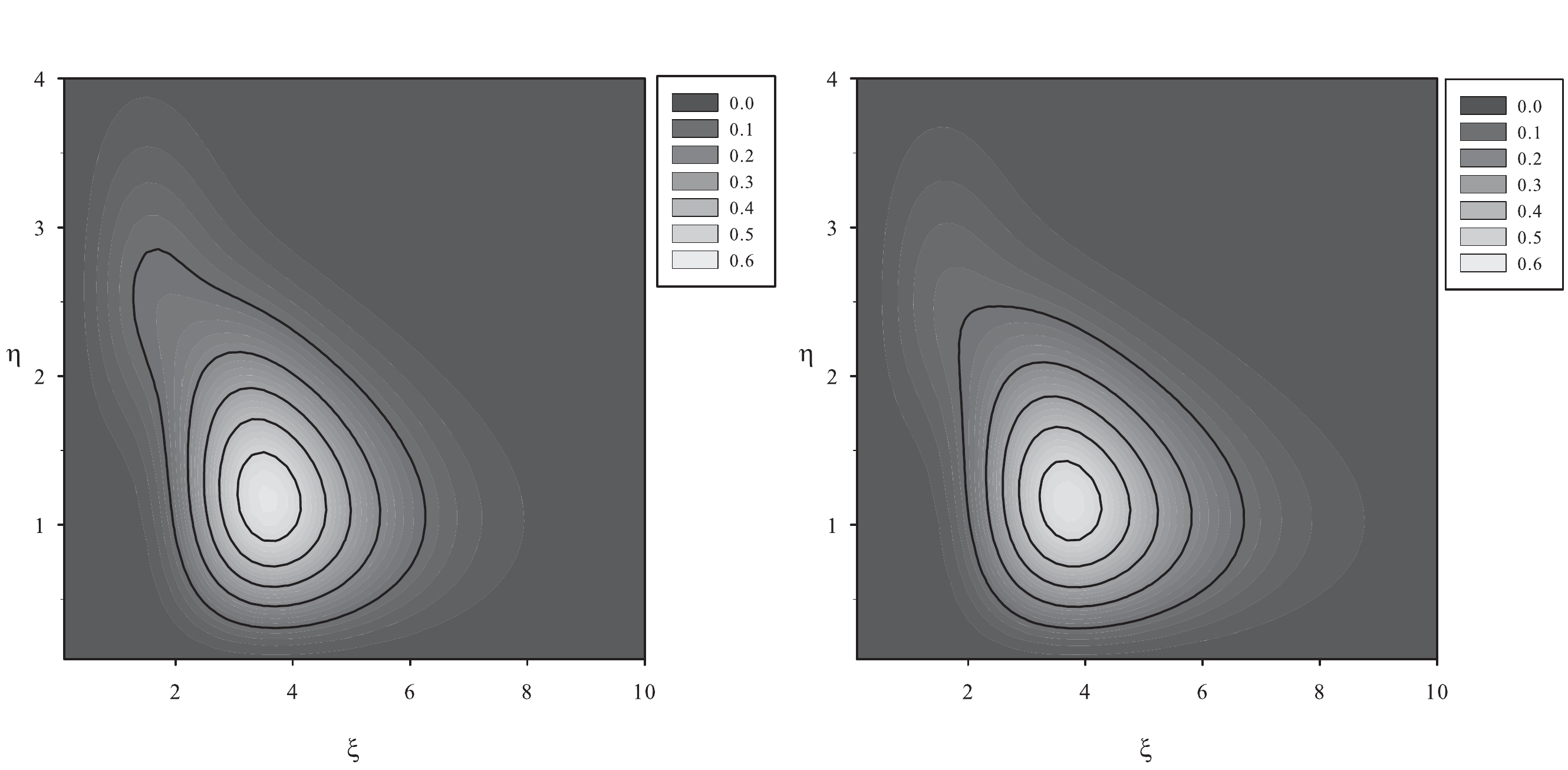}
\end{center}
\caption{Phase portraits of the $3/2^{-}$ (left) and $1/2^{-}$ (right) bound
states in $^{7}Be$.}
\label{Fig:PP_BStates7Be}
\end{figure}

\subsection{Phase portrait for resonance states.}

\begin{figure}[ht]
\begin{center}
\includegraphics[width=\textwidth]{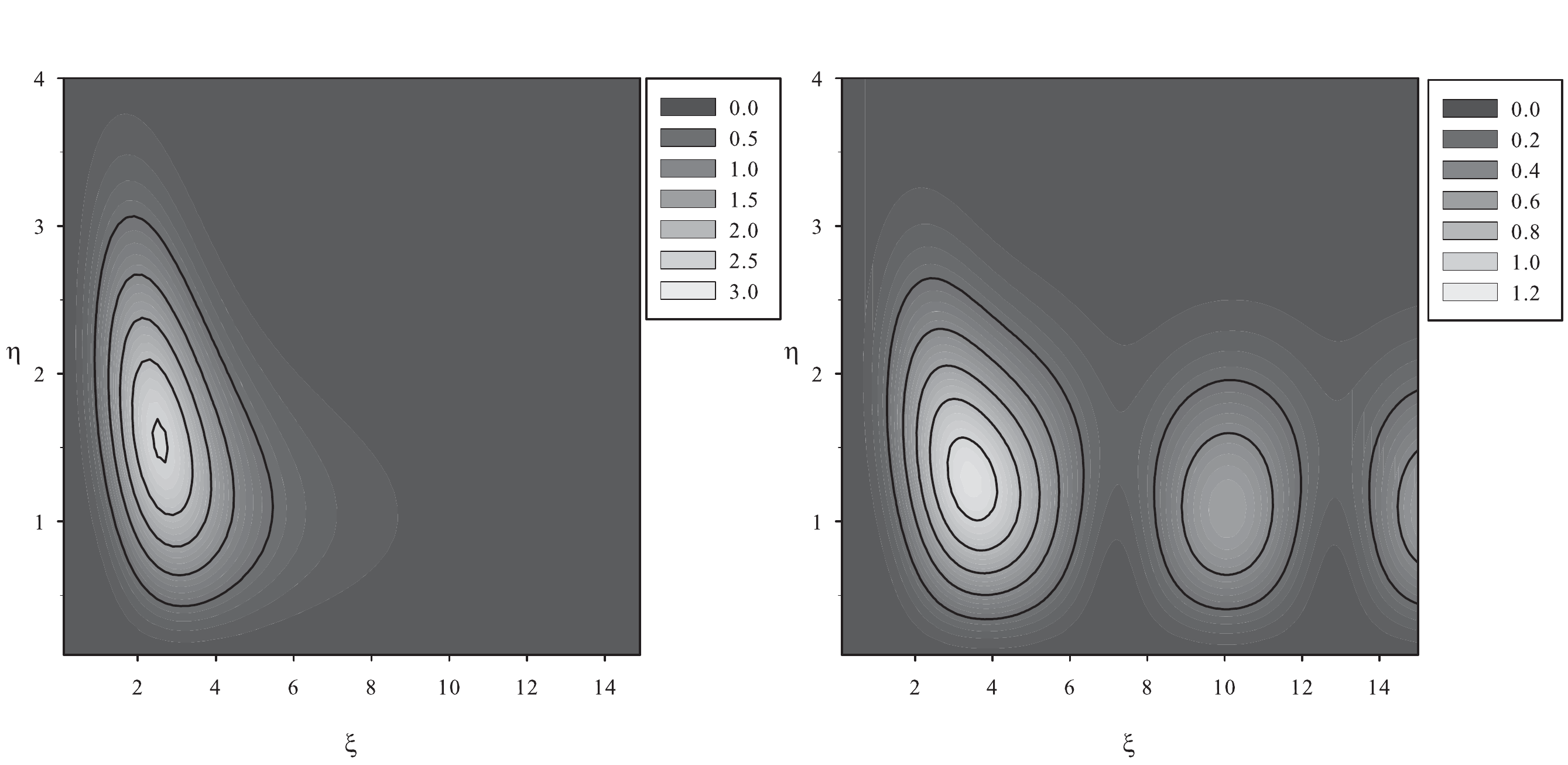}
\end{center}
\caption{Phase portraits of the $3^{+}$ (left) and $2^{+}$ (right) resonance
states in $^{6}Li$.}
\label{Fig:PP_Resons6Li}
\end{figure}

\begin{figure}[ptbh]
\begin{center}
\includegraphics[width=\textwidth]{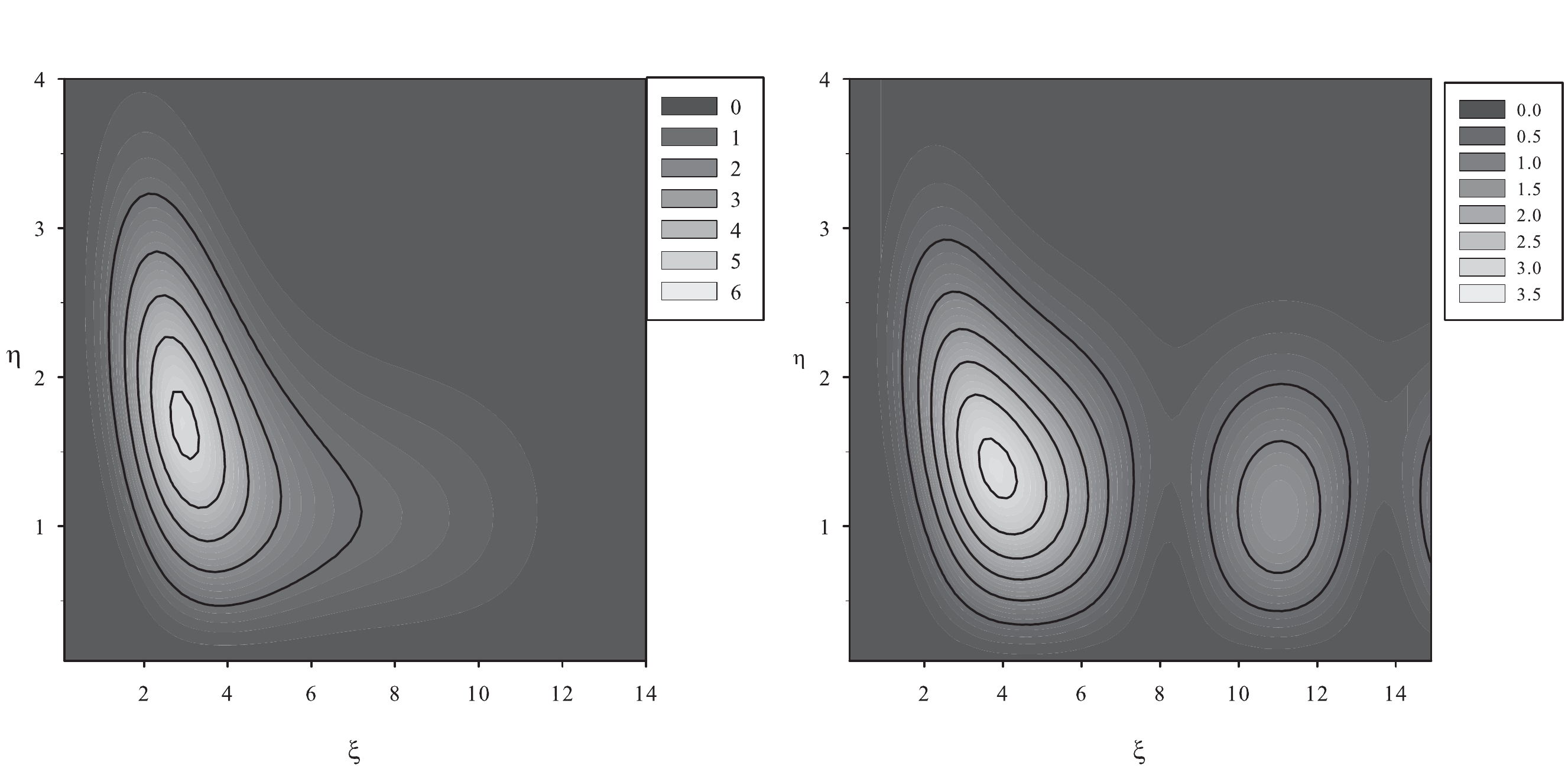}
\end{center}
\caption{Phase portraits of the $\frac{7}{2}^{-}$ and $\frac{5}{2}^{-}$
resonance states in $^{7}Li$.}
\label{Fig:PP_Resons7Li}
\end{figure}

\begin{figure}[ptbh]
\begin{center}
\includegraphics[width=\textwidth]{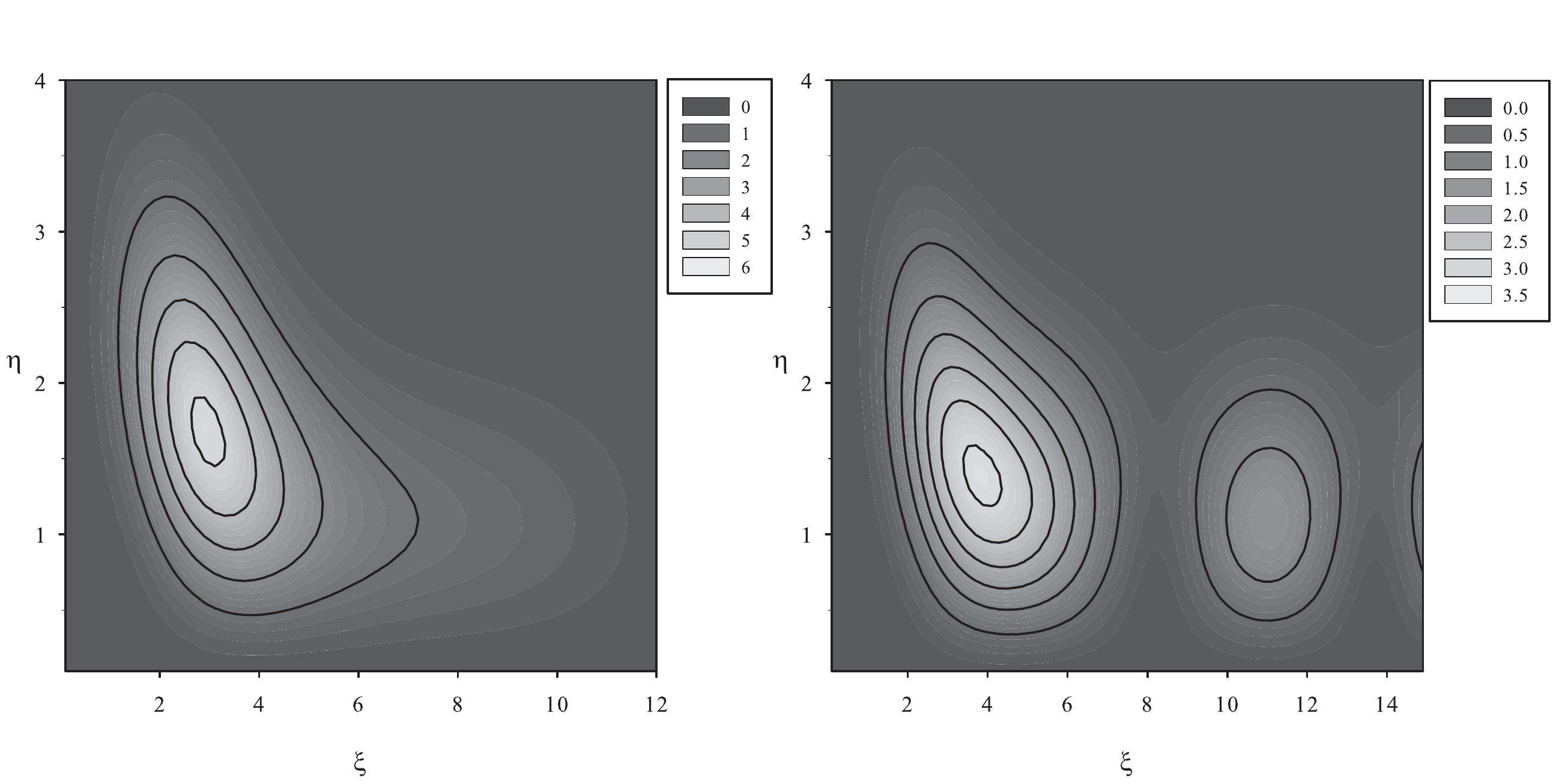}
\end{center}
\caption{Phase portraits of the $\frac{7}{2}^{-}$ and $\frac{5}{2}^{-}$
resonance states in $^{7}Be$.}
\label{Fig:PP_Resons7Be}
\end{figure}

There are narrow and broad resonance states in the nuclei of interest. The
narrowest resonance state is observed in $^{8}Be$. Indeed, the calculated
width of the $0^{+}$ resonance state is only 2.40 eV. The broadest resonance
state is also observed in $^{8}Be$. The width of the $4^{+}$ resonance states
exceeds 3.5 MeV. One would expect quite different density distributions and
shape of phase portraits for narrow and broad resonance states. This is so,
as can be seen from Figures \ref{Fig:PP_Resons6Li}, \ref{Fig:PP_Resons7Li}, %
\ref{Fig:PP_Resons7Be}, where we display the phase portraits for the resonance
states in $^{6}Li$, $^{7}Li$ and $^{7}Be$. Each nucleus is represented by
two resonance states, one of which is narrow ($3^{+}$ in $^{6}Li$, $7/2^{-}$
in $^{7}Li$ and $^{7}Be$) \ and the other is broad ($2^{+}$ in $^{6}Li$, $%
5/2^{-}$ in $^{7}Li $ and $^{7}Be$). When we are saying "a narrow resonance
state" or "a broad resonance state", we mean not only the absolute value of
the resonance width, but also the value of $\Gamma/E$. For narrow resonance
states this ratio is below 0.067, and for broad resonance states it exceeds
0.34. \ For the narrowest $0^{+}$ resonance state in $^{8}Be$, the ratio $%
\Gamma/E\approx3\cdot10^{-5}$.

There is a strong resemblance between narrow resonance states and bound
states. Both of them realize themselves in a compact area of the phase space.

As for the broad resonance states, they have a principal maximum at relatively
small values of coordinates $\xi$ and momenta $\eta$. Besides, they also have many
regular maxima at a fixed value of $\eta$, but different values of $\xi$.
This structure reflects an oscillatory behavior of the coordinate wave function
and $\delta$-like behavior of the wave function in the momentum space.

\begin{figure}[ptbh]
\begin{center}
\includegraphics[width=\textwidth]{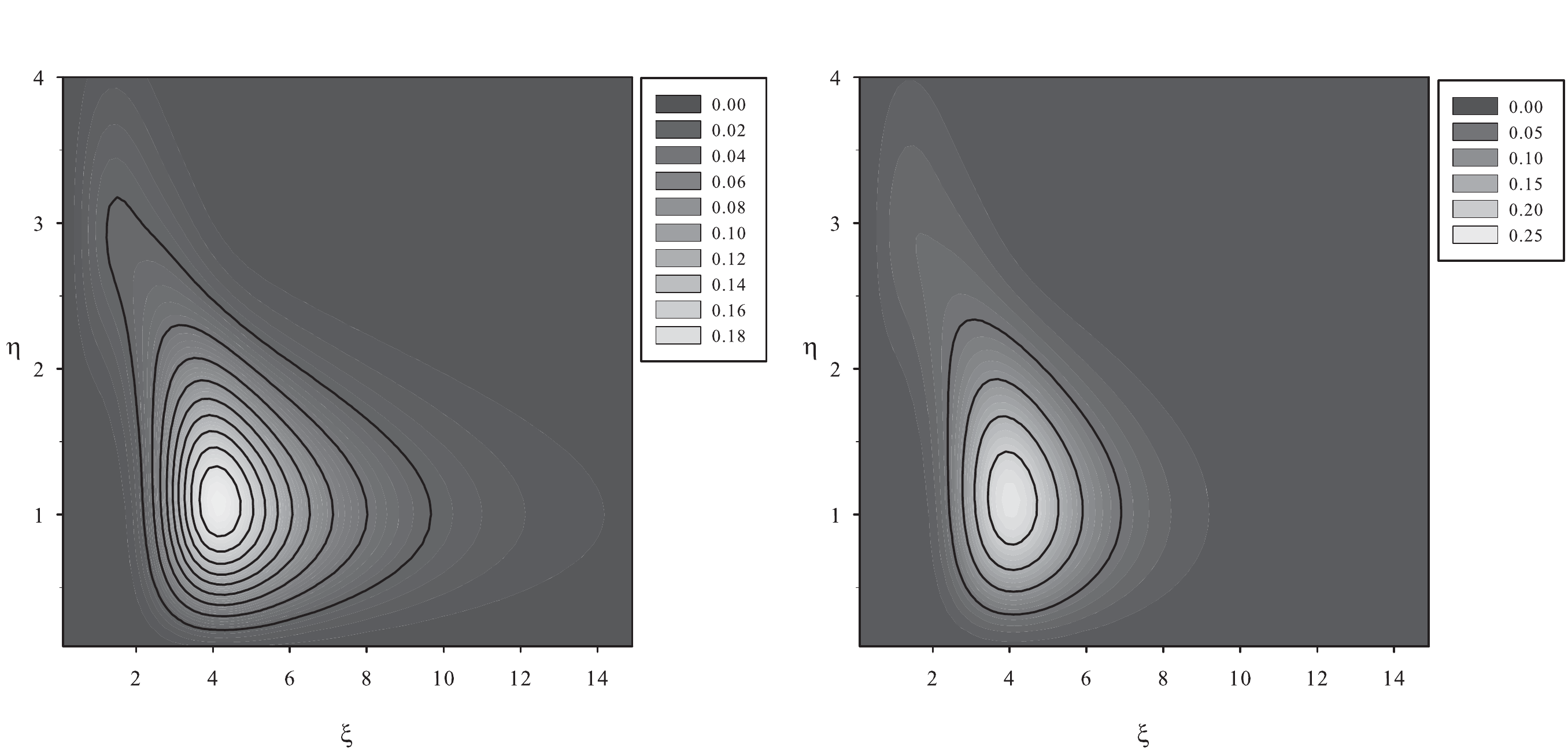}
\end{center}
\caption{Phase portraits of $0^{+}$ resonance (left) and bound (right)
states in $^{8}Be$. }
\label{Fig:PP_RS_BS_0P8Be}
\end{figure}

In Figure \ref{Fig:PP_RS_BS_0P8Be} we compare phase portraits of the $0^{+}$
"ground" state in $^{8}Be$, obtained with and without the Coulomb
interaction (RS and BS calculations). The energy of the resonance state is
0.0818 MeV above the $\alpha +\alpha$ threshold, while the energy of the
bound state is -1.3529 MeV bellow the threshold.

One can see, that the phase portraits of resonance and bound states are quite
similar. This similarity can be attributed to the fact that in the internal
region, where interaction between clusters is strong, wave functions have
very close behavior. The asymptotic part of these functions are totally
different. For a bound state, the function $\psi \left( q\right) $
exponentially decreases \ (\ref{eq:C20C}), while it slowly decreases (\ref%
{eq:C21C}) for a resonance state. Thus, the asymptotic part of wave
functions gives a small contribution to a density distribution in the
Fock-Bargmann space. To prove that in the internal region wave function of
resonance and bound state are similar, we display \ in Figures \ref%
{Fig:DD_CS_MS_0P_8Be} the density distributions in the coordinate and momentum
representations. Indeed, these figures show that there is small difference
between the wave functions or density distributions of the bound and resonance
states in $^{8}Be$. Maximum of density distributions is approximately in the
same point of the coordinate space. However, maxima of density distribution of
a BS\ and RS in the momentum space lie at different values of momentum $p$, which
connected with different energies of the resonance state (0.0809 MeV) and bound
state (-1.3529 MeV). Since the wave function of the bound state is more compact in
the coordinate space than the resonance one, it is more dispersed in the momentum
space.

\begin{figure}[ptbh]
\begin{center}
\includegraphics[width=\textwidth]{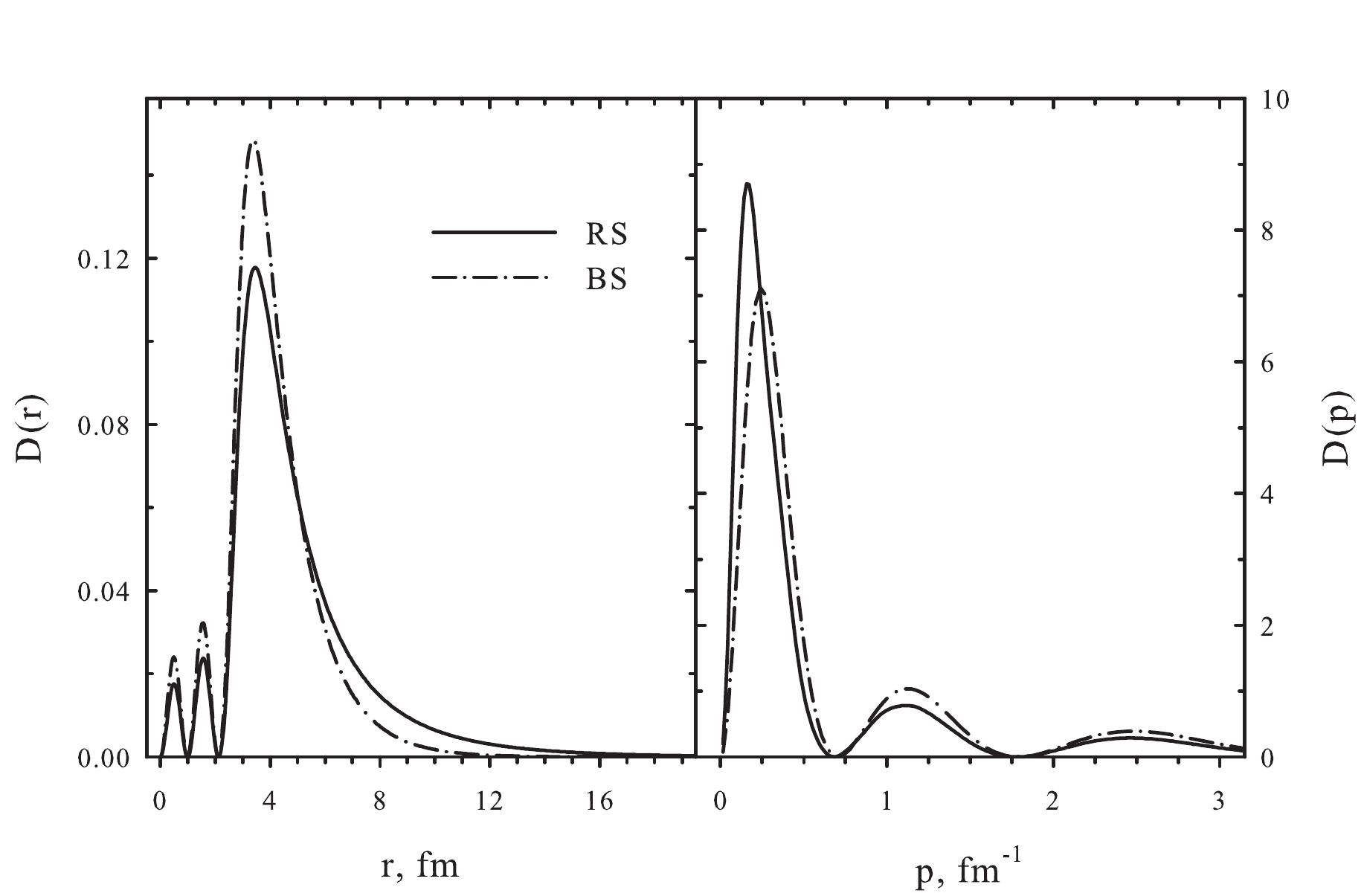}
\end{center}
\caption{Density distributions of the $0^{+}$ resonance and bound states in $%
^{8}Be$ in coordinate (left) and momentum (right) spaces.}
\label{Fig:DD_CS_MS_0P_8Be}
\end{figure}

It is interesting to compare a density distribution in the phase space with
density distributions in the coordinate and momentum spaces. \ For this aim, we
find a point in the phase space ($\xi_{\max}$, $\eta_{\max}$) such that the density
distributions has maximum. And then, we calculate the density distributions $%
D\left( \xi,\eta _{\max}\right) $\ and $D\left( \xi_{\max},\eta\right) $.
Thus from a large number of trajectories, we selected only two of them, the first
one is \ with a fixed momentum $\eta_{\max}$, the second one is with a fixed
coordinate $\xi$. The first trajectory we compare with the density
distribution in the coordinate space $D\left( r\right) $, and the second
trajectory is compared with the density distribution in the momentum space $%
D\left( p\right) $. These results are presented in Figure \ref%
{Fig:PPvsDD_Reson0P8Be}. One can see, that the density distributions $D\left(
\xi,\eta_{\max}\right) $\ and $D\left( r\right) $ are quite close to each
other, they have maximum approximately at the same point $\xi \simeq r$.
There are more differences in behavior of the density distributions $D\left(
\xi_{\max},\eta\right) $\ and $D\left( p\right) $. Maximum of $D\left(
p\right) $ is shifted to smaller values of $p$, as compared to the function $%
D\left( \xi_{\max},\eta\right) $. Such relation between $D\left(
\xi,\eta_{\max}\right) $\ and $D\left( r\right) $, $D\left(
\xi_{\max},\eta\right) $\ and $D\left( p\right) $ is also observed for bound
and resonance states of other nuclei.

\begin{figure}[ptbh]
\begin{center}
\includegraphics[width=\textwidth]{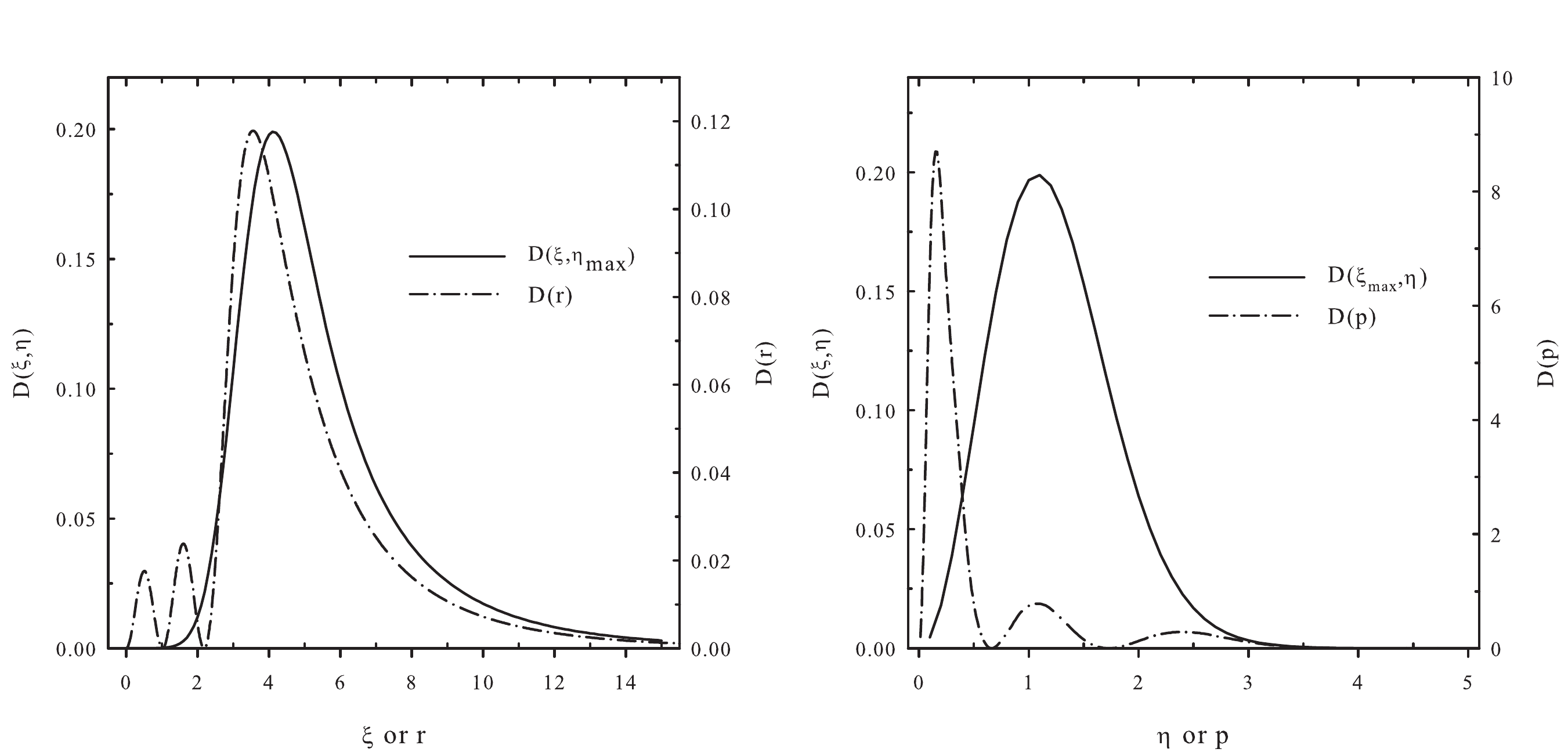}
\end{center}
\caption{Density distributions in phase space and in coordinate and momentum
representations for $0^{+}$ resonance state in $^{8}Be$. See more detail in
text.}
\label{Fig:PPvsDD_Reson0P8Be}
\end{figure}

\begin{figure}[ptbh]
\begin{center}
\includegraphics[width=\textwidth]{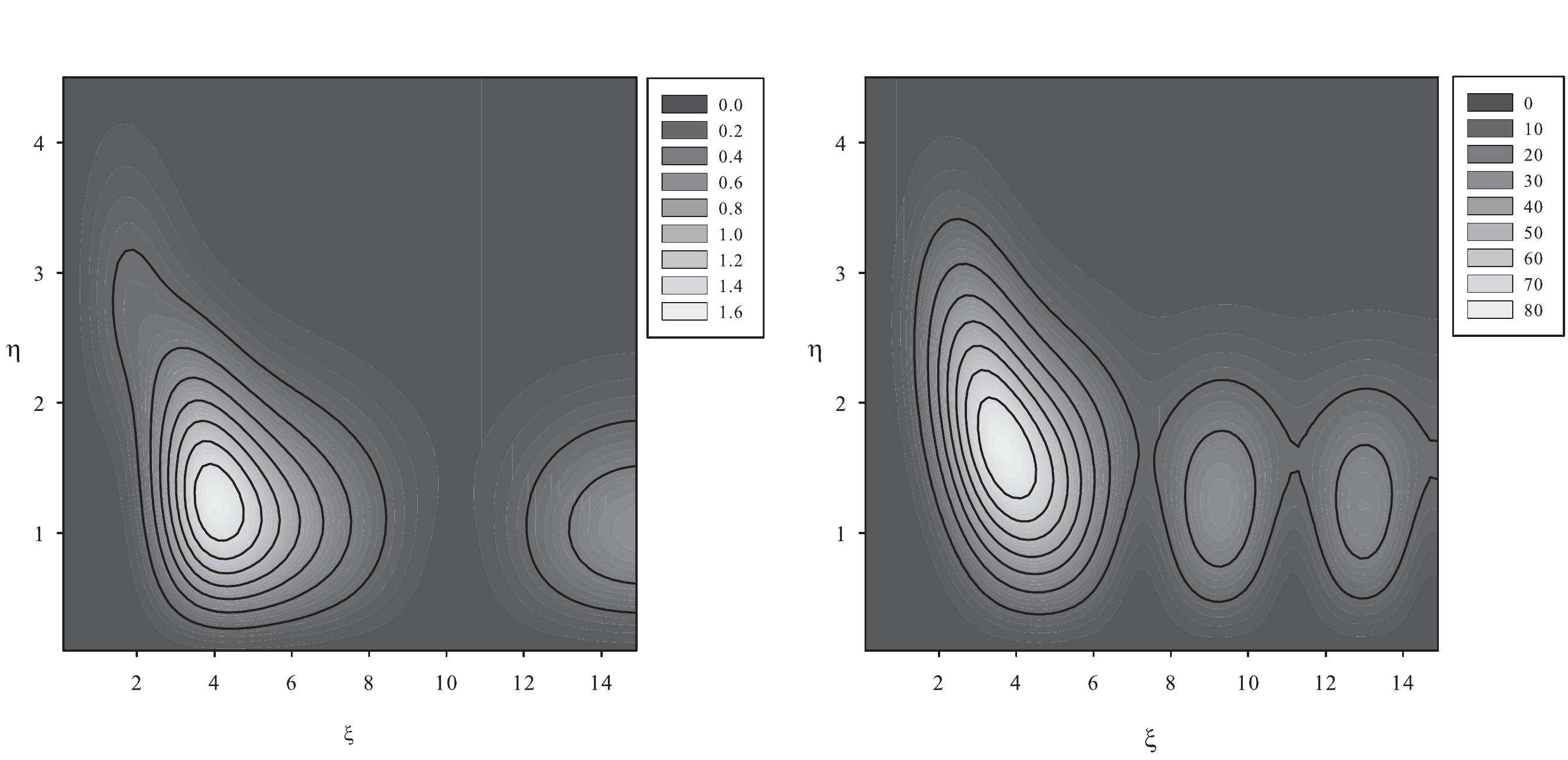}
\end{center}
\caption{Phase portraits for the $2^{+}$ (left) and $4^{+}$ (right)
resonance states in $^{8}Be$.}
\label{Fig:PP_Resons8Be}
\end{figure}

By closing this section, in Figure \ref{Fig:PP_Resons8Be} we demonstrate
the phase portraits for the broad $2^{+}$ and $4^{+}$ resonance states. One can see
that the oscillations of the density distribution along the axis $\xi$ is more
frequent for the $4^{+}$ resonance state than for the $2^{+}$ resonance state. This
is expected feature of the density distribution, because the energy of the $4^{+}$
resonance state exceeds the energy of the $2^{+}$ resonance state by a
factor of 8. Both resonance states reveal strong quantum effects, because
they have prominent maximum in the region of small intercluster distances in the
phase space.

As with the bound states, a maximum of the density distribution in phase space
for the resonance states is also observed at the values of variable $\eta$,
which significantly differ from the corresponding value of the dimensionless
parameter $k$. To demonstrate this, we consider the resonance states in $^{8}Be$%
. The dimensionless parameter $k$ for $^{8}Be$, where the $0^{+}$ resonance
state has the smallest energy and the $4^{+}$ resonance state has the largest
energy, varies at the range 0.086$\leq k\leq$0.944. However, a maximum of the
density distribution for the $4^{+}$ resonance state is achieved at $%
\eta\approx1.5$. This value has to be compared with $k=$0.944. For the $%
0^{+} $ resonance state, the momentum $k$=0.086; and the density distribution peaks
at $\eta\approx1.1$.

\subsection{Classical regime}

In this subsection, we consider the high energy excited state of the two-cluster
systems. So far we considered excited states with the energy less than 10 MeV.
Now we will look for the range of energy, where phase portraits have an evident
"classical shape" or where all quantum trajectories approach classical
trajectories.

\begin{figure}[ptbh]
\begin{center}
\includegraphics[width=\textwidth]{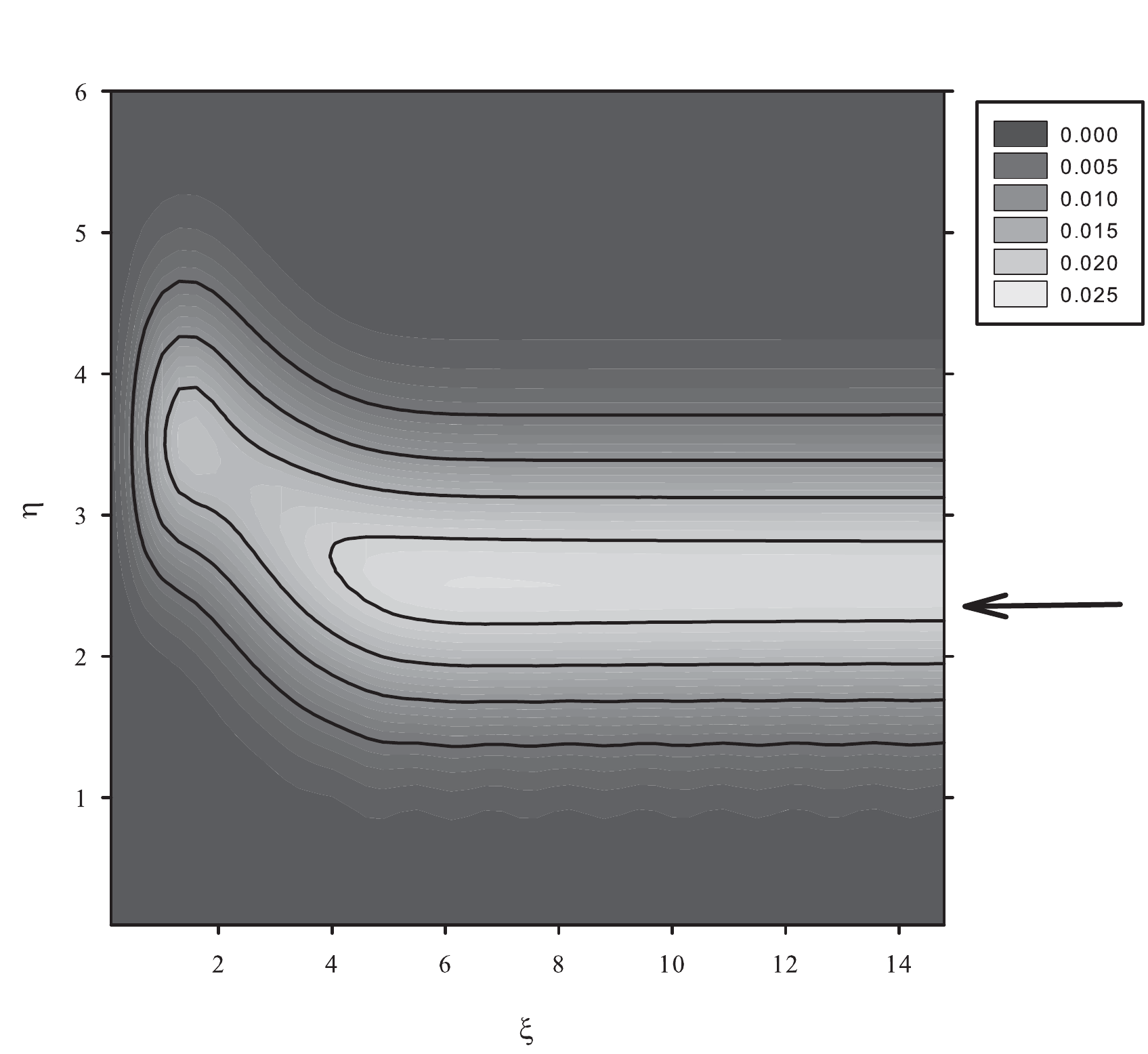}
\end{center}
\caption{Phase portrait for $0^{+}$ excited state in $^{8}Be$ with energy $E$%
=60.21 MeV and dimensionless momentum $k$=2.34 (this value is indicated by
arrow).}
\label{Fig:PP_0P8BeE60}
\end{figure}

Let us consider the excited $0^{+}$ states of $^{8}Be$ with energy $E$=60.21
MeV, Figure \ref{Fig:PP_0P8BeE60}. This energy corresponds to the
dimensionless momentum $k=b\sqrt{\frac{2mE}{\hbar ^{2}}}$=2.34. One can see
that the maximum of the density distribution is observed for such value of $%
\eta $ which is very close to the corresponding value of momentum $k$. This
means that the classical regime is realized for a relatively small value of the
excitation energy of the two-cluster system. Besides, one can see that the density
distribution, presented in Figure \ref{Fig:PP_0P8BeE60}, is similar to that
for free motion of a particle with a large value of the momentum $k$ (cf. with Fig. \ref%
{eq:03} and eq. (\ref{dro_plane})). It is important to underline,
that the classical regime becomes valid for a moderate value of
the coordinate $\xi $.

The correspondence between a quantum density distribution and its classical limit
is observed for other values of the total angular momenta in the $^{8}Be$ and
for all angular momenta in other nuclei as well. The larger is the energy of the
excited state, the closer is the density distribution to the classical
trajectory.

\section{Conclusions \label{Sect:Conclusion}}

We have studied the trajectories of two-cluster nuclear systems in phase
space. The Bargmann-Segal transformation has been used to map wave functions
of two-cluster systems in the coordinate space into the Fock-Bargmann space. \
The two-cluster systems have been studied within a microscopic model which makes
use of a full set of oscillator functions to expand a wave function of
relative motion of interacting clusters. The dominant two-cluster partition
of each nuclei was taken into consideration. The input parameters of
the model and nucleon-nucleon potential were selected to optimize
description of the internal structure of clusters and to reproduce position of
the ground state with respect to the two-cluster threshold. We have considered
a wide range of excitation energies of compound systems, but special attention
was devoted to the bound and resonance states. It was shown that bound
states and narrow resonance states realize themselves in a very compact area
of the phase space. Phase portraits of the excited states with large value of the energy
have a maximal value along the line which coincides with a classical trajectory.

We have also considered two model problems, a harmonic oscillator and free motion in a three-dimensional case, which helped us to understand the dynamics of real physical systems in the phase space.

\section{Acknowledgment}
This work was supported in part by the Program of Fundamental
Research of the Physics and Astronomy Department of the National
Academy of Sciences of Ukraine.

\bibliographystyle{elsarticle-num}




\begin{thebibliography}{10}
\expandafter\ifx\csname url\endcsname\relax
  \def\url#1{\texttt{#1}}\fi
\expandafter\ifx\csname urlprefix\endcsname\relax\def\urlprefix{URL }\fi
\expandafter\ifx\csname href\endcsname\relax
  \def\href#1#2{#2} \def\path#1{#1}\fi

\bibitem{1986JPSJ...55..762T}
K.~{Takahashi}, {Wigner and Husimi Functions in Quantum Mechanics}, J. Phys.
  Soc. Jpn. 55 (1986) 762.
\newblock \href {http://dx.doi.org/10.1143/JPSJ.55.762}
  {\path{doi:10.1143/JPSJ.55.762}}.

\bibitem{PhaseSpace2005}
{C. K. Zachos, D. B. Fairlie, and T. L. Curtright}, Quantum Mechanics in Phase
  Space, World Scientific, Singapore, 2005.

\bibitem{1999JMP....40.2531M}
K.~B. {M{\o}ller}, {Comment on phase-space representation of quantum state
  vectors}, J. Math. Phys. 40 (1999) 2531--2535.
\newblock \href {http://dx.doi.org/10.1063/1.532881}
  {\path{doi:10.1063/1.532881}}.

\bibitem{1993JChPh..98.3103T}
G.~{Torres-Vega}, J.~H. {Frederick}, {A quantum mechanical representation in
  phase space}, J. Chem. Phys. 98 (1993) 3103--3120.
\newblock \href {http://dx.doi.org/10.1063/1.464085}
  {\path{doi:10.1063/1.464085}}.

\bibitem{1997JChPh.106.7228M}
K.~B. {M{\o}ller}, T.~G. {J{\o}rgensen}, G.~{Torres-Vega}, {On coherent-state
  representations of quantum mechanics: Wave mechanics in phase space}, J.
  Chem. Phys. 106 (1997) 7228--7240.
\newblock \href {http://dx.doi.org/10.1063/1.473684}
  {\path{doi:10.1063/1.473684}}.

\bibitem{1990JChPh..93.8862T}
G.~{Torres-Vega}, J.~H. {Frederick}, {Quantum mechanics in phase space: New
  approaches to the correspondence principle}, J. Chem. Phys. 93 (1990)
  8862--8874.
\newblock \href {http://dx.doi.org/10.1063/1.459225}
  {\path{doi:10.1063/1.459225}}.

\bibitem{1928ZPhy...49..339F}
V.~{Fock}, {Verallgemeinerung und L{\"o}sung der Diracschen statistischen
  Gleichung}, Zeit. Phys. 49 (1928) 339--357.
\newblock \href {http://dx.doi.org/10.1007/BF01337923}
  {\path{doi:10.1007/BF01337923}}.

\bibitem{Bargmann:1946me}
V.~Bargmann, {Irreducible unitary representations of the Lorentz group}, Ann.
  Math. 48 (1947) 568--640.
\newblock \href {http://dx.doi.org/10.2307/1969129}
  {\path{doi:10.2307/1969129}}.

\bibitem{Perelomov_GCS}
A.~Perelomov, Generalized coherent states and their applications, Berlin,
  Springer, 1986.

\bibitem{2003CRPhy...4..497K}
Y.~{Kanada-En'yo}, M.~{Kimura}, H.~{Horiuchi}, {Antisymmetrized Molecular
  Dynamics: a new insight into the structure of nuclei}, C. R. Physique 4
  (2003) 497--520.
\newblock \href {http://dx.doi.org/10.1016/S1631-0705(03)00062-8}
  {\path{doi:10.1016/S1631-0705(03)00062-8}}.

\bibitem{2012PTEP.2012aA202K}
Y.~{Kanada-En'yo}, M.~{Kimura}, A.~{Ono}, {Antisymmetrized molecular dynamics
  and its applications to cluster phenomena}, Prog. Theor. Exp. Phys. 2012~(1)
  (2012) 010000.
\newblock \href {http://arxiv.org/abs/1202.1864} {\path{arXiv:1202.1864}},
  \href {http://dx.doi.org/10.1093/ptep/pts001}
  {\path{doi:10.1093/ptep/pts001}}.

\bibitem{2000RvMP...72..655F}
H.~{Feldmeier}, J.~{Schnack}, {Molecular dynamics for fermions}, Rev. Mod.
  Phys. 72 (2000) 655--688.
\newblock \href {http://arxiv.org/abs/cond-mat/0001207}
  {\path{arXiv:cond-mat/0001207}}, \href
  {http://dx.doi.org/10.1103/RevModPhys.72.655}
  {\path{doi:10.1103/RevModPhys.72.655}}.

\bibitem{2004NuPhA.738..357N}
T.~{Neff}, H.~{Feldmeier}, {Cluster structures within Fermionic Molecular
  Dynamics}, Nucl. Phys. A 738 (2004) 357--361.
\newblock \href {http://arxiv.org/abs/arXiv:nucl-th/0312130}
  {\path{arXiv:arXiv:nucl-th/0312130}}, \href
  {http://dx.doi.org/10.1016/j.nuclphysa.2004.04.061}
  {\path{doi:10.1016/j.nuclphysa.2004.04.061}}.

\bibitem{1937PhRv...52.1083W}
J.~A. {Wheeler}, {Molecular Viewpoints in Nuclear Structure}, Phys. Rev. 52
  (1937) 1083--1106.
\newblock \href {http://dx.doi.org/10.1103/PhysRev.52.1083}
  {\path{doi:10.1103/PhysRev.52.1083}}.

\bibitem{2005PPN..36.714F}
G.~{Filippov}, Y.~{Lashko}, {Structure of Light Neutron-Rich Nuclei and Nuclear
  Reactions Involving These Nuclei}, Phys. Part. Nucl. 36~(6) (2005) 714--739.

\bibitem{2013FBS.55.817L}
Y.~A. {Lashko}, G.~F. {Filippov}, V.~S. {Vasilevsky}, M.~D. {Soloha-Krymchak},
  {Phase Portraits of Quantum Systems}, Few-Body Syst. 55~(8-10) (2014)
  817--820.
\newblock \href {http://dx.doi.org/10.1007/s00601-013-0760-8}
  {\path{doi:10.1007/s00601-013-0760-8}}.

\bibitem{kn:wilderm_eng}
K.~Wildermuth, Y.~Tang, A unified theory of the nucleus, Vieweg Verlag,
  Braunschweig, 1977.

\bibitem{kn:Fil_Okhr}
G.~F. Filippov, I.~P. Okhrimenko, Use of an oscillator basis for solving
  continuum problems, Sov. J. Nucl. Phys. {\bf 32} (1981) 480--484.

\bibitem{kn:Fil81}
G.~F. Filippov, On taking into account correct asymptotic behavior in
  oscillator-basis expansions, Sov. J. Nucl. Phys. {\bf 33} (1981) 488--489.

\bibitem{kn:cohstate2E}
G.~F. Filippov, V.~S. Vasilevsky, L.~L. Chopovsky, Solution of problems in the
  microscopic theory of the nucleus using the technique of generalized coherent
  states, Sov. J. Part. and Nucl. {\bf 16} (1985) 153--177.

\bibitem{kn:Minn_pot1}
D.~R. Thompson, M.~LeMere, Y.~C. Tang, Systematic investigation of scattering
  problems with the resonating-group method, Nucl. Phys. {\bf A286}~(1) (1977)
  53--66.
\newblock \href {http://dx.doi.org/10.1016/0375-9474(77)90007-0}
  {\path{doi:10.1016/0375-9474(77)90007-0}}.

\bibitem{1970NuPhA.158..529R}
I.~{Reichstein}, Y.~C. {Tang}, {Study of N + {$\alpha$} system with the
  resonating-group method}, Nucl. Phys. A 158 (1970) 529--545.
\newblock \href {http://dx.doi.org/10.1016/0375-9474(70)90201-0}
  {\path{doi:10.1016/0375-9474(70)90201-0}}.

\bibitem{2002NuPhA.708....3T}
D.~R. {Tilley}, C.~M. {Cheves}, J.~L. {Godwin}, G.~M. {Hale}, H.~M. {Hofmann},
  J.~H. {Kelley}, C.~G. {Sheu}, H.~R. {Weller}, {Energy levels of light nuclei
  \makebox{$A$}=5, 6, 7}, Nucl. Phys. A 708 (2002) 3--163.
\newblock \href {http://dx.doi.org/10.1016/S0375-9474(02)00597-3}
  {\path{doi:10.1016/S0375-9474(02)00597-3}}.

\bibitem{2004NuPhA.745..155T}
D.~R. {Tilley}, J.~H. {Kelley}, J.~L. {Godwin}, D.~J. {Millener}, J.~E.
  {Purcell}, C.~G. {Sheu}, H.~R. {Weller}, {Energy levels of light nuclei
  {$A$}=8, 9, 10}, Nucl. Phys. A 745 (2004) 155--362.
\newblock \href {http://dx.doi.org/10.1016/j.nuclphysa.2004.09.059}
  {\path{doi:10.1016/j.nuclphysa.2004.09.059}}.

\bibitem{Saito69}
S.~Saito, {Interaction between Clusters and Pauli Principle}, Prog. Theor.
  Phys. {\bf 41}~(3) (1969) 705--722.
\newblock \href {http://dx.doi.org/10.1143/PTP.41.705}
  {\path{doi:10.1143/PTP.41.705}}.

\bibitem{kn:Saito77}
S.~Saito, {Theory of Resonating Group Method and Generator Coordinate Method,
  and Orthogonality Condition Model}, Prog. Theor. Phys. Suppl. {\bf 62} (1977)
  11--89.
\newblock \href {http://dx.doi.org/10.1143/PTPS.62.11}
  {\path{doi:10.1143/PTPS.62.11}}.

\end{thebibliography}


\end{document}